\documentclass[aps,prx,reprint,showpacs,superscriptaddress,longbibliography]{revtex4-2}
\usepackage{mathtools,amssymb,graphicx,units}

\usepackage[plainpages=false,pdfpagelabels,colorlinks=true,linkcolor=red,urlcolor=blue,citecolor=blue,pdftitle={Title},pdfauthor={},pdfdisplaydoctitle=true,pdfduplex=DuplexFlipLongEdge]{hyperref}
\usepackage{verbatim}
\usepackage[english]{babel}
\usepackage{xcolor}
\usepackage{dsfont}

\newcommand{\tcc}{\textcolor{black}}

\usepackage[normalem]{ulem}

\newcommand{\beginsupplement}{%
        \setcounter{table}{0}
        \renewcommand{\thetable}{S\arabic{table}}%
        \setcounter{figure}{0}
        \renewcommand{\thefigure}{S\arabic{figure}}%
}

\begin{document}

\title{Universality and Critical Exponents of the Fermion Sign Problem}
\author{R. Mondaini}
\email{rmondaini@csrc.ac.cn}
\author{S. Tarat}
\email{tarats@csrc.ac.cn}
\affiliation{Beijing Computational Science Research Center, Beijing 100193, China}
\author{R.T. Scalettar}
\email{scalettar@physics.ucdavis.edu}
\affiliation{Department of Physics and Astronomy, University of California,
Davis, CA 95616, USA}
 
\begin{abstract}
Initial characterizations of the fermion sign problem focused on its evolution with spatial lattice size $L$ and inverse temperature $\beta$, emphasizing the implications of the exponential nature of the decay of the average sign $\langle {\cal S} \rangle$ for the complexity of its solution and associated limitations of quantum Monte Carlo studies of strongly correlated materials. Early interest was also on the evolution of $\langle {\cal S} \rangle$  with density $\rho$,  either because commensurate filling is often associated with special symmetries for which the sign problem is absent, or because particular fillings are often primary targets, e.g.~those densities which  maximize superconducting transition temperature (the top of the `dome' of cuprate systems). Here we describe a new analysis of the sign problem which demonstrates that the {\it spin-resolved} sign $\langle {\cal S}_\sigma\rangle$ already possesses signatures of universal behavior traditionally associated with order parameters, even in the absence of symmetry protection that makes $\langle {\cal S} \rangle = 1$. When appropriately scaled, $\langle {\cal S}_\sigma \rangle$ exhibits universal crossings and data collapse. Moreover, we show these behaviors occur in the vicinity  of quantum critical points of three well-understood models, exhibiting either second-order or Kosterlitz-Thouless phase transitions. Our results pave the way for using the average sign as a minimal correlator that can potentially describe quantum criticality in a variety of fermionic many-body problems.
\end{abstract}

\maketitle
\section{Introduction}
The sign problem is the fundamental obstacle that prevents accurate computations in a variety of problems of quantum correlated matter. In evading the `exponential wall' that precludes the application of unbiased methods, such as exact diagonalization~\cite{Poilblanc1991,Lin1993} and matrix-product-states-based algorithms~\cite{Schollwock2005}, for large systems or arbitrary dimensions, quantum Monte Carlo techniques~\cite{Loh1992,gubernatis16} have in principle the potential to solve fundamental questions, including understanding $d$-wave pairing mechanisms of repulsive fermions, for example~\cite{Dagotto1994,Qin2021}. Yet, the fact that the importance sampling of quantum configurations is not constrained to render positive weights severely limits its applicability in the most salient class of quantum problems of interest.

Since generic solutions are not always available, a common approach relies on restricting computations to regimes where the sign problem is still well-behaved, allowing the extraction of statistically  convergent quantities. Recent developments based on finding a local basis that mitigates it~\cite{Marvian2019,Hangleiter2020,Levy2021}, or fine-tuned Hubbard-Stratonovich transformations to delay its onset~\cite{Wan2020}, have been very useful but do not provide an overarching circumvention scheme. Although solutions to this conjectured NP-hard problem~\cite{Troyer2005} are unlikely to be discovered anytime soon, a precise investigation of the onset of the sign problem, and its relation to the physics of the underlying studied Hamiltonian, is much less explored and potentially of great impact. Further investigations have advanced the idea that this connection does exist, and in different models, the appearance of the sign problem seems coupled to the manifestation of quantum critical behavior~\cite{mondaini2021,Wessel2017}.

When several non-hybridizing fermionic (spin) species are present, the sign problem involves the product of contributions from each component. Previous studies have focused on this product, both because it is required to weigh physical observables and also because, in several important cases, the product is better behaved than its constituents.  Furthermore, existing work has generally concentrated on the `scaling behavior'  in the sense of large space-time systems, i.e., how the sign evolves as the inverse temperature $\beta \rightarrow \infty$ and spatial size $L \rightarrow \infty$.

In this manuscript, we introduce two novel aspects of the study of the sign problem and show that they constitute a powerful new approach to using quantum simulations to explore many-body physics. First, we analyze the `spin-resolved sign' and argue that the usual approach of examining the average product of the sign of the individual weights can actually obscure physical content inherent when spin-resolution is used. Second, we examine the behavior of the sign near phase transitions, both those which occur as quantum critical points, through the variation of a parameter in the Hamiltonian and thermal phase transitions, which occur as temperature $T$ is lowered. 
 
Taken together, we demonstrate that the spin-resolved sign can be used to locate phase transitions and determine critical exponents. Furthermore, it has the potential to do so {\it even more accurately} than traditional observables ${\cal A}$ such as spin, charge, and pairing correlations. The reason is that these latter quantities require a precise measurement of {\it ratios} $\langle {\cal A} {\cal S} \rangle / \langle {\cal S} \rangle$ of quantities with increasing fluctuations originating both inherently in the physics (response functions are themselves measurements of fluctuations) and in the vanishing of the sign.  The (spin-resolved) sign, by itself, is thus a  less noisy `observable' if it can be shown, as we do here, that it holds information about criticality. 

In what follows, we first investigate three fermionic models showing numerically that a scaling analysis of the average weights aids in the characterization of either known quantum or thermal phase transitions. We then provide a \textit{demonstration of why} this happens, i.e., we provide a theoretical justification for
our observation that the average weights display indicators of criticality. We also further include information in support of the dynamic critical exponent used in finite-temperature calculations to promote scaling.

\section{The SU(2) Honeycomb Hubbard model}
We initially investigate the spinful Hubbard model on a honeycomb lattice with $N=2L^2$ sites, 
\begin{equation}
\hat H = 
-t \sum_{\langle ij\rangle \,\sigma}  \hat c^{\dagger}_{i\sigma} \hat c^{\phantom{\dagger}}_{j\sigma} 
+ U \sum_{i} \hat n_{i\uparrow} \hat n_{i\downarrow} 
-\mu\sum_{i,\sigma}\hat n_{i\sigma},
\label{eq:ham_spinful}
\end{equation}
where $\hat c^{\dagger}_{i\sigma}$ ($\hat c^{\phantom\dagger}_{i\sigma}$) creates (annihilates) a fermion at site $i$ with spin $\sigma$, and $\hat n_{i\sigma}$ is the number density operator. With an increasing magnitude of the ratio of the amplitude of the interaction to the hopping scale, $U/t$, the ground-state at half-filling (chemical potential $\mu = U/2$) exhibits a continuous phase transition from a Dirac semi-metal to a Mott insulator featuring antiferromagnetic order. This transition, described by an effective quantum field-theory model (Gross-Neveu)~\cite{Gross1974}, belongs to the chiral Heisenberg universality class, and has been characterized in numerics in a variety of fermionic lattice models~\cite{Sorella2012,Assaad2013,Toldin2015,Otsuka2016,Zhang2019,Otsuka2020}.

High precision computation of the critical interaction in \eqref{eq:ham_spinful} yields $U_c/t = 3.78-3.87$, with critical exponent in the range $\nu = 0.84-1.02$~\cite{Sorella2012,Assaad2013,Toldin2015,Otsuka2016} associated to the divergence of the correlation length in the vicinity of the critical point $\xi \propto |U-U_c|^{-\nu}$. These values are obtained via the scaling of physical observables: the staggered magnetization order parameter~\cite{Sorella2012,Assaad2013,Otsuka2016}, single particle gap~\cite{Assaad2013}, and quasiparticle weight~\cite{Otsuka2016,Otsuka2020}. Here, instead, we propose an analysis based on the average sign. Although a non-physical observable, tied to the computational method used, the average sign is, however, required to compute \textit{any} physical observable in a quantum Monte Carlo (QMC) simulation. Our results thus suggest that the sign problem is inextricably linked to the determination of the physics of the model. In the following two sections, we explore generic aspects of the sign problem and how they apply to this particular Hamiltonian. Subsequently, we build on this knowledge to understand quantum and thermal phase transitions in other fermionic models.

\section{The sign problem}
We start by recalling a known scaling form of the average sign in QMC calculations. It originates from considering the definition in terms of the weights $W$ of the configurations $\{x\}$ sampled in $D$ spatial and one imaginary-time dimension as~\cite{Loh1990,Troyer2005},
\begin{equation}
    \langle {\cal S}\rangle = \frac{\sum_{\{x\}} W(\{x\})}{\sum_{\{x\}} |W({\{x\}})|} = \frac{{\cal Z}_W}{{\cal Z}_{|W|}}.
\end{equation}
Here, $W(\{x\}) = \det M_\uparrow(\{x\})\cdot \det M_\downarrow(\{x\})$ is a product of weights of individual fermionic flavors in the case of Eq.~\eqref{eq:ham_spinful} for determinant QMC calculations (see Appendix~\ref{app:Methods} for specific definitions in the various models we study and Appendix~\ref{app:Matrix} for an analysis of the matrices $M_\sigma$); ${\cal Z}_{W}$ is the partition function of the original problem in its formulation in $D+1$ dimensions~\cite{Blankenbecler1981,Hirsch1985}, whereas ${\cal Z}_{|W|}$ instead uses the positive-definite absolute value of the weight to proceed with the importance sampling in the simulations. Written in terms of the  corresponding free energy densities, $f=[-1/(\beta N)]\log {\cal Z}$, the average sign thus reduces to $\langle {\cal S}\rangle  = \exp[-\beta N (f_W - f_{|W|})]$. Given that $\sum_{\{x\}} W(\{x\}) \leq \sum_{\{x\}} |W({\{x\}})|$ and that the free energy is extensive, it follows that $f_W \geq f_{|W|}$, and thus the average sign exponentially decreases in terms of both real-space and imaginary-time dimensions~\cite{Loh1990,Iglovikov2015}, if not protected by some symmetry of the problem~\cite{Wu2005,Wei2016}.

\begin{figure}[t]
\centering
\includegraphics[width=0.99\columnwidth]{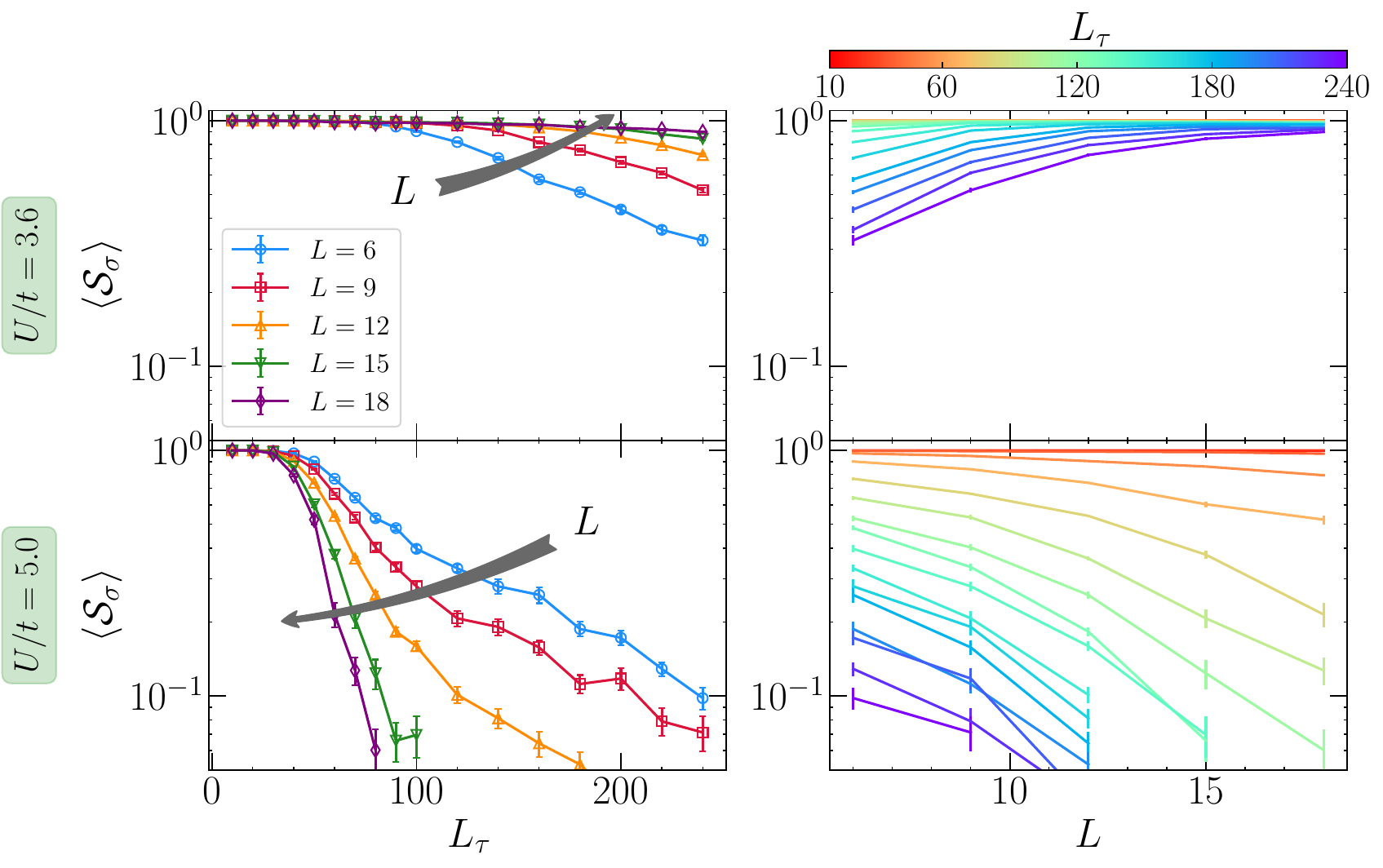}
\caption{Dependence of the average spin resolved sign on the space-imaginary time dimensions. $\langle {\cal S}_\sigma\rangle$ vs.~$L_\tau$ (left) and $L$ (right) for the SU(2) honeycomb Hubbard model, before ($U/t = 3.6$, top) and after ($U/t = 5.0$, bottom) the putative transition $U_c$. For $U<U_c$ ($U>U_c$), the average spin resolved sign \textit{grows} (\textit{reduces}) with increasing lattice size at low temperatures. Here and elsewhere, error bars denote the standard error of the mean of independent realizations. The imaginary-time discretization is $\Delta\tau \equiv \beta/L_\tau = 1/10$.}
\label{fig:scaling_L_tau_L}
\end{figure}

\begin{figure*}[t]
\centering
\includegraphics[width=0.94\textwidth]{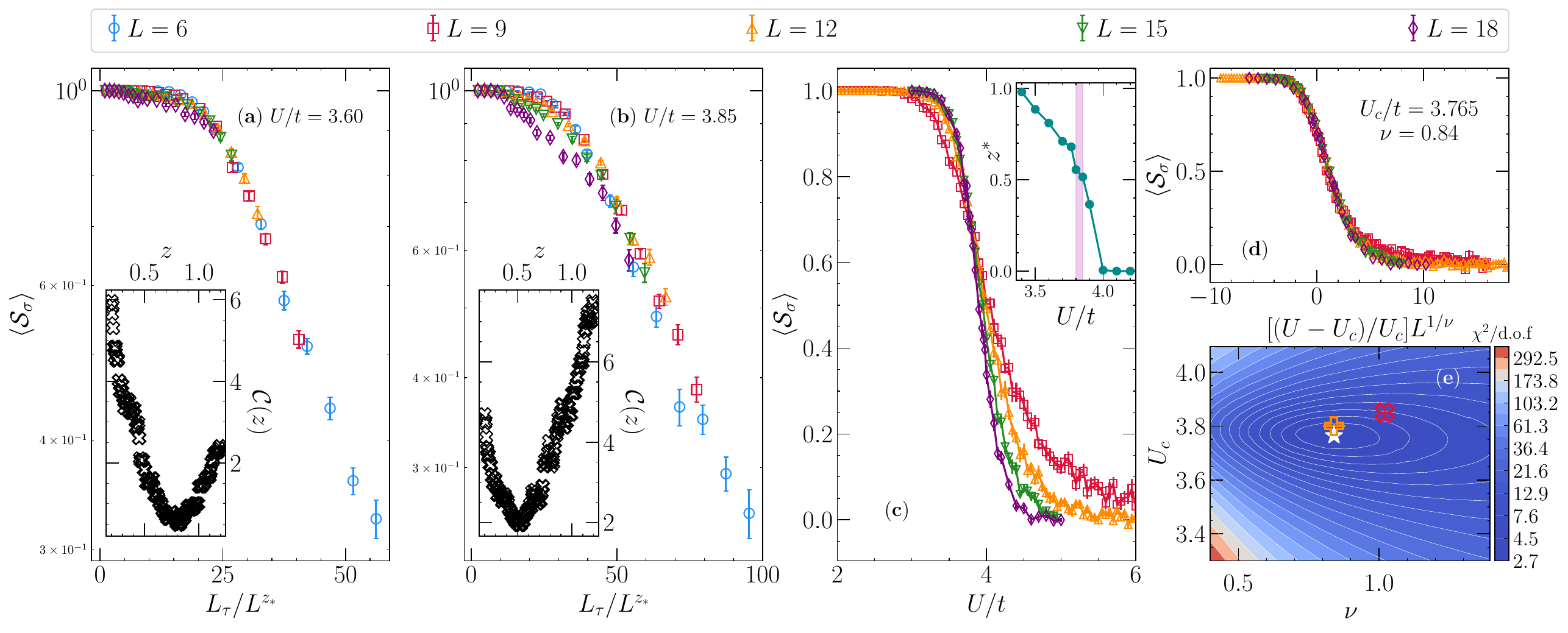}
\caption{Scaling analysis of the spin-resolved sign in the honeycomb SU(2) Hubbard model. Scaling in the vicinity of the best-known estimations of the critical point $U_c/t$ [(a) $U/t=3.6$ and (b) $U/t=3.85$] using a re-scaled x-axis $L_\tau/L^z$. The insets display a cost function that determines the collapse quality of $\langle {\cal S}_\sigma\rangle(L_\tau,L)$ at different values of the dynamic critical exponents $z$ (see text). A compilation is given in the inset of (c) at a range of $U/t$ values; estimations for the critical interactions from Refs.~\cite{Toldin2015,Otsuka2016} are marked by the shaded region. (c) $\langle {\cal S}_\sigma\rangle$ \textit{vs.} $U/t$ and different lattice sizes at half-filling; the number of imaginary time-slices used roughly preserves the ratio $L_\tau/L^z$ fixed, $L_\tau = 240$, 220, 196, 170 for $L=18,$ 15, 12, 9, respectively; that is we use $z\simeq0.5$. (d) Scaling using a functional form $g[uL^{1/\nu}]$ whose critical exponent $\nu$, as obtained by minimizing the error $\chi^2/{\rm d.o.f}$ of a high-order polynomial fitting in the space of parameters $(U_c,\nu)$. (e) The contour plot of $\chi^2/{\rm d.o.f}$, where the minimum is at $U_c/t=3.765$ and $\nu=0.84$ as shown by the star symbol. Recent estimations using physical quantities for the same model~\cite{Toldin2015,Otsuka2016} are annotated by the cross markers. Here, $t\, \Delta\tau=0.1$ is used.
}
\label{fig:scaling_hc}
\end{figure*}

An example of this protection is the case of Eq.~\eqref{eq:ham_spinful} at half-filling. Via a $\downarrow$-spin particle-hole transformation, $c_{i\downarrow} = (-1)^i c_{i\downarrow}^\dag$, where $(-1)^i=+1(-1)$ on the $A(B)$ sublattice of the bipartite honeycomb geometry (or any other bipartite lattice), the weight simplifies to ${\rm const.} \times [\det M_\uparrow(\{x\})]^2 $ for whichever configuration $\{x\}$, when using a spin-decomposed Hubbard-Stratonovich transformation~\cite{Hirsch1983,Hirsch1985}. In general, however, symmetries that preclude the onset of the sign problem are not available for most models of interest. 

\section{The spin-resolved sign}
Although the `total' sign problem has been investigated in detail~\cite{Loh1990,Iglovikov2015}, the properties of the sign of individual determinants that compose the weight in models with a larger number of local degrees of freedom were much less explored. Moreover, past work focused on the behavior as $\beta\to\infty$ and not near the critical point. By systematically computing the average sign of the determinant of a \textit{single} spin-species, $\langle {\cal S}_\sigma\rangle \equiv \sum_{\{x\}} {\rm sgn}(\det M_\sigma \{ x\}) |W(\{ x\})|/\sum_{\{x\}} |W(\{ x\})|$, for the problem in Eq.~\eqref{eq:ham_spinful}, we have earlier demonstrated~\cite{mondaini2021} (see corresponding Supplementary Materials) a behavior reminiscent of an order parameter undergoing a typical phase transition: it displays its maximum value $\langle {\cal S}_\sigma\rangle = 1$ in the quantum disordered phase while $\langle {\cal S}_\sigma\rangle \to 0$ in the ordered region ($U>U_c$) at sufficiently low temperatures [see Fig.~\ref{fig:scaling_hc}(c), for example]. The latter occurs in spite of the fact that $\langle {\cal S}\rangle$ is pinned at one since we take $\mu = U/2$, dictating thus that in the ordered regime the most likely configurations $\{x\}$ display random signs of $\det M_\sigma(\{x\})$.

This behavior, including a crossing of the curves for different system sizes at $U\simeq U_c$, is suggestive that a scaling function $g$, for the spin-resolved sign exists, similar in motivation to those used for traditional, physical observables to characterize quantum criticality:
\begin{equation}
    \langle {\cal S}_\sigma\rangle(u,L,L_\tau) = g(u L^{1/\nu}, L_\tau/L^z),
    \label{eq:scal_func}
\end{equation}
where $L$ is the linear system size, $L_\tau = \beta/\Delta\tau$ is the number of imaginary-time slices of the inverse temperature in writing down the path integral ${\cal Z}$, $u = (U-U_c)/U_c$ is the reduced coupling, and $z$ is the dynamic critical exponent. The second argument comes from the fact that the $D+1$ lattice is anisotropic in its dimensions (and eventual effective couplings)~\cite{Binder1989,Rieger1994,Guo1994}. The previous empirical observation determined that for $u>0$, $g(x,y)\to0$ when both $x,y$ diverge.

To better understand the limits of $\langle {\cal S}_\sigma\rangle$, we display in Fig.~\ref{fig:scaling_L_tau_L} its dependence on both $L_\tau$ and $L$. We notice that the previous expectation that the average total sign exponentially decreases with the system size is also valid for $\langle {\cal S}_\sigma\rangle$, but provided that $U\gtrsim U_c$ and temperatures are sufficiently low ($L_\tau \gg L$). If in the quantum disordered phase, however, the spin-resolved sign \textit{increases} with growing $N$. As we will see, this contrasting behavior is fundamental for the identification of the critical interactions using the average sign of individual determinants.

\begin{figure*}[htp!]
\centering
\includegraphics[width=0.94\textwidth]{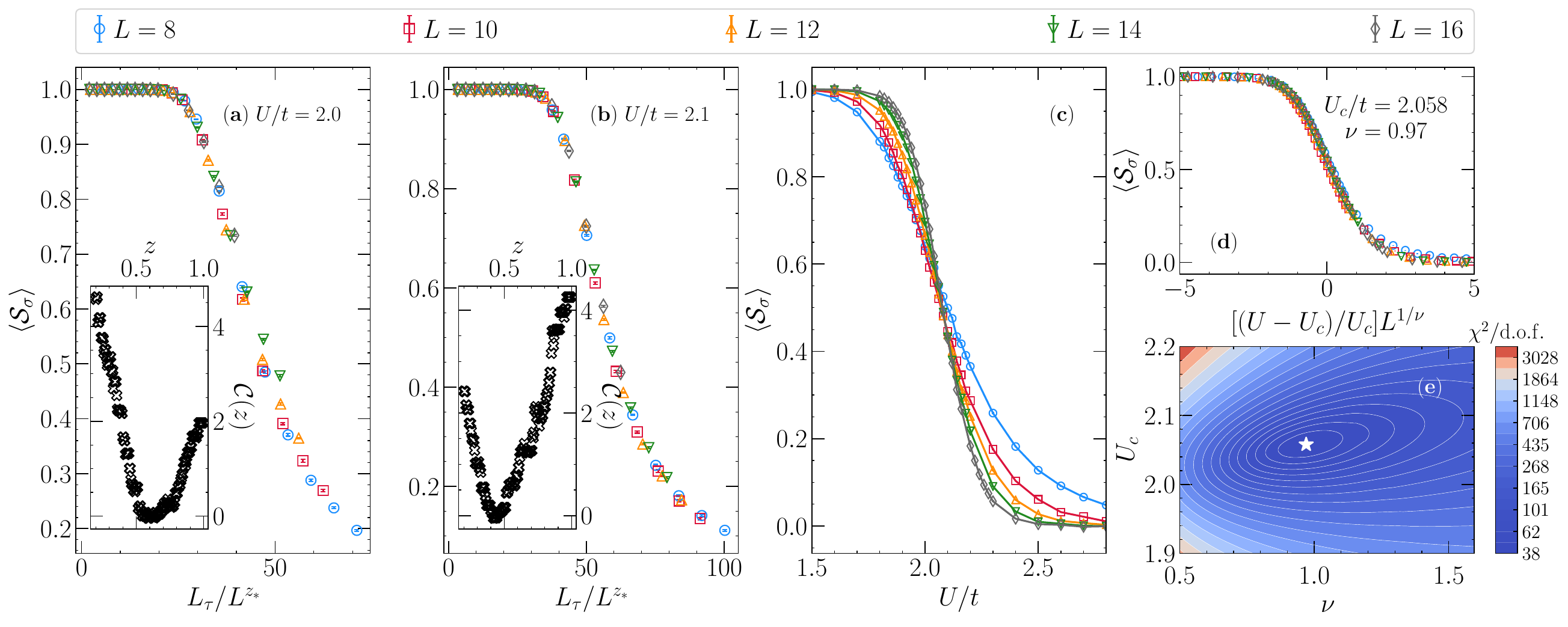}
\caption{Scaling analysis of the spin-resolved sign in the square lattice SU(2) Ionic Hubbard. (a,b) Similar to Fig.~\ref{fig:scaling_hc} (a,b) but for the SU(2) Ionic Hubbard model on the square lattice, for values of $U/t = 2$ (a) and 2.1 (b). (c) $\langle {\cal S}_\sigma\rangle$ \textit{vs.} $U/t$ and different lattice sizes at half-filling; the number of imaginary time slices is chosen such that $L_\tau/L^{0.5} \simeq 50$. (d) Scaling of the data in  (c) using a functional form $g[uL^{1/\nu}]$, whose critical exponent $\nu=0.97$ and critical interaction $U_c/t=2.058$ are extracted from an analysis (e) as done in Fig.~\ref{fig:scaling_hc}(e). All results are obtained at $\Delta/t=0.5$; other values lead to similar results but with different $U_c/t$ critical values. Here, we use $t\, \Delta\tau=0.1$.
}
\label{fig:ionic}
\end{figure*}

At first sight, owing to the known Lorentz invariance that emerges at the critical point~\cite{Herbut2009,Herbut2009b}, the dynamic critical exponent is surmised as $z=1$. Yet, we do not take this as a starting point, relaxing this assumption to show that a smaller value in $\langle {\cal S}_\sigma\rangle$ actually gives an optimal scaling. Motivation for $z\neq1$ is provided in  Sec.~\ref{sec:z}. We thus try to scale $\langle {\cal S}_\sigma\rangle$ in the vicinity of the critical point with a functional form $L_\tau/L^z$, as displayed in Figs.~\ref{fig:scaling_hc}(a) and \ref{fig:scaling_hc}(b), a procedure which has been argued to improve the scaling of related quantum models~\cite{Rieger1994,Guo1994}. By defining a cost function ${\cal C}(z) = \sum_j (|y_{j+1} - y_j|)/(\max\{y_j\} - \min\{y_j\})-1$~\cite{Suntajs2020,Aramthottil2021}, where $y_j$ are the values of $\langle {\cal S}_\sigma\rangle(L_\tau,L)$, ordered according to their $L_\tau/L^z$ ratio, the dynamic critical exponent $z^*$ that minimizes ${\cal C}$ can be extracted, see insets in Figs.~\ref{fig:scaling_hc}(a) and (b). A compilation of the $z^*(U)$ values is given in the inset of Fig.~\ref{fig:scaling_hc}(c), accompanied by a range of recently known predictions of $U_c$~\cite{Toldin2015,Otsuka2016}. It is clear that close to the critical point, the scaling with the second argument of the function $g$ should be taken with $z\simeq 1/2$, for the current range of imaginary-time slices $L_\tau$ (or inverse temperatures $\beta$ with $t\Delta\tau=0.1$) used.

Hence, we use this current estimation to proceed with scaling in order to simultaneously obtain the critical exponent $\nu$ and the critical interaction $U_c$. With the $L_\tau/L^z$ ratio fixed, a clear crossing of the average sign of individual weights when increasing the lattice size can be seen [Fig.~\ref{fig:scaling_hc}(c)], accurately determining the critical interaction. By using the functional form of Eq.~\eqref{eq:scal_func}, we obtain the collapse of the average spin-resolved sign [Fig.~\ref{fig:scaling_hc}(d)], yielding a critical  exponent $\nu\simeq0.84$ and $U_c/t \simeq 3.77$. This estimation, obtained by minimizing the error of a high-order polynomial fit to the data in the space of parameters $(U,\nu)$, is shown in Fig.~\ref{fig:scaling_hc}(e) [See the Supplemental Materials (SM)~\cite{SM} for a different method of scaling analysis]. One can contrast these results with recent estimations using the same model, as in Ref.~\onlinecite{Toldin2015} with $U_c/t = 3.80(1)$ and $\nu = 0.84(4)$, while in Ref.~\onlinecite{Otsuka2016}, $U_c/t = 3.85(1)$ and $\nu = 1.02(1)$, both using a zero-temperature version of the QMC method employed here~\cite{Loh1992,assaad02}. While larger system sizes and other finite corrections may improve our results, they are already in quite remarkable agreement with the best estimations to date.

\section{The SU(2) Ionic Hubbard model}
The preceding discussion provided \textit{quantitative} evidence that the average sign of a single determinant contains precise information about the quantum criticality in a well-studied model; it remains an open question whether this is general. Here we provide compelling further validation by looking at one of the simplest models that bypass the symmetry that prevents the onset of the sign problem, the ionic Hubbard model on the square lattice~\cite{fabrizio99,kampf03,garg06,paris07,bouadim07,craco08,garg14}. That is, in a model that in the standard fermionic basis suffers from the sign problem even at half-filling. Here $\hat H_{\rm Ionic} = \hat H + \Delta\sum_{i\sigma} (-1)^i \hat n_{i,\sigma}$, adds a staggered onsite potential proportional to $\Delta$ to the Hamiltonian~\eqref{eq:ham_spinful}, which we investigate again at the average density of one electron per site. 

\begin{figure*}[htp!]
\centering
\includegraphics[width=1\textwidth]{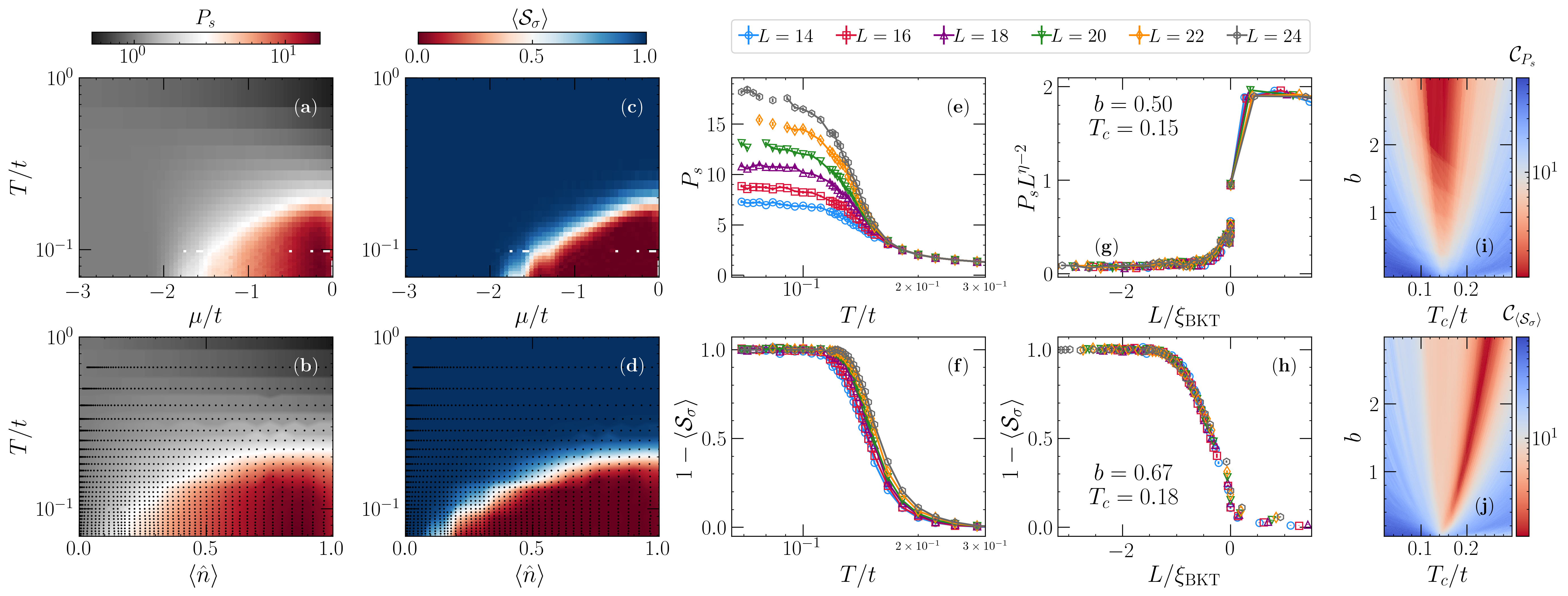}
\caption{KT-scaling analysis of the SU(2) square lattice attractive Hubbard model. The color plot of the $s$-wave pair structure factor $P_s$ (a) and the spin-resolved sign $\langle {\cal S}_\sigma\rangle$ (b) in the $T$ vs.~$\mu$ plane on a $L=16$ lattice; (c) and (d) display the same if using the extracted average density $\langle \hat n\rangle$ instead, with black markers depicting the outputs on the regular grid in $\mu$. (e) and (f) show a cut along the $\mu/t=-0.75$ line for $P_s$ and $\langle {\cal S}_\sigma\rangle$ and different lattice sizes. (g) and (h) use a KT-scaling form to collapse the curves with (i) and (j) displaying the corresponding cost function ${\cal C}$ of the scaling in the $(T_c,b)$ parameter space. The KT-scaling is performed such that $L/\xi_{\rm BKT}$ for $T>T_c$ and $-L/\xi_{\rm BKT}$ if $T<T_c$. All data are extracted at $U/t=-4$ and $t\Delta\tau = 1/16$.
}
\label{fig:attractive}
\end{figure*}

The {\it qualitative} physics of this model at finite $\Delta$ is generally agreed to display a competition of band insulating ($\Delta\gg U$), Mott insulating ($U \gg \Delta$), and metallic behavior when both interactions and staggered potential magnitudes are comparable~\cite{paris07,bouadim07}. A recent investigation \cite{mondaini2021} has indicated that this correlated metal phase can be qualitatively tracked by the regime where the average  sign of the QMC weights vanishes. We now employ our new sign scaling method to understand the critical behavior at the transition from the  band-insulator to the metallic phase, the quantitative details of which are still  under debate in the community.

We fix the staggered potential at $\Delta/t = 0.5$, while increasing the interactions $U$ to overcome the externally imposed (i.e.~by the one-body potential in $\hat H$) charge density wave induced by $\Delta$. As before, we estimate the value of the dynamic critical exponent in the vicinity of the transition, $U_c (\Delta=0.5)/t \simeq 2.0$~\cite{paris07,bouadim07} in Fig.~\ref{fig:ionic} (a,b) by scaling  the $\langle {\cal S}_\sigma\rangle$ with $L_\tau/L^z$, resulting in $z\simeq0.5$. Using thus a roughly fixed ratio $L_\tau/L^{0.5}$, we provide the scaling of the spin-resolved sign in Fig.~\ref{fig:ionic}(c,d). Here fluctuations are small, and the scaling renders an accurate determination of the critical interactions driving the band-insulator-to-metal transition $U_c/t \simeq 2.05$ with related critical exponent $\nu = 0.97$. 

The one-dimensional version of this model has been extensively studied via numerics~\cite{Lou2003,Manmana2004,Tincani2009}, and a field-theory close to the quantum critical points exists~\cite{Fabrizio1999}. The transition where the band-insulating phase ends, with its externally imposed charge density wave giving way to a dimerized bond-ordered wave insulator, belongs to the 2D Ising universality class in that case. Here in the two-dimensional model, QMC results point out to a band-insulator to correlated metallic transition~\cite{paris07,bouadim07}, whose universality class is unknown and where field-theories describing it are currently not available, precluding a direct comparison of the calculated exponent $\nu$, obtained from the scaling of $\langle {\cal S}_\sigma\rangle$, with existing knowledge.

\section{The attractive Hubbard model}
We now generalize these two results for quantum critical behavior in the ground-state to finite-temperature transitions. A well-studied example is the onset of superconductivity in the two-dimensional negative-$U$ SU(2) Hubbard model: for chemical potentials $\mu \neq |U|/2$, there is a Kosterlitz-Thouless (KT) transition at temperature $T_c\neq 0$~\cite{Scalettar1989, Moreo1991, Paiva2004, Fontenele2022} to a superconducting phase. As a direct consequence of the often used (charge-decomposed) Hubbard-Stratonovich transformation~\cite{Hirsch1983}, the weight matrices $M_\sigma$ are identical, resulting in the complete absence of sign problem since the remaining single-particle part of the Hamiltonian is equal for both spin species. That the total sign is always positive does not preclude that the average sign of \textit{individual} weights converge to zero; this can be seen in Fig.~\ref{fig:attractive}(c), which shows this quantity in the $T$ vs.~$\mu$ parameter space on  an $L=16$ square lattice. The regime $\langle {\cal S}_\sigma\rangle \to 0$ is directly related to the one where the $s$-wave equal-time pair structure factor $P_s=(1/L^2)\sum_{i,j} \langle \hat \Delta_i \hat \Delta_j^\dagger\rangle$ ($\hat \Delta_i \equiv \hat c_{i\uparrow} \hat c_{j\downarrow}$) is also large [Fig.~\ref{fig:attractive}(a)].

To make this comparison quantitative, Figs.~\ref{fig:attractive}(e) and \ref{fig:attractive}(f) display a finite-size analysis at a fixed $\mu/t=-0.75$, which gives densities close to $\langle \hat n\rangle=0.5$ at low temperatures~\cite{Scalettar1989,SM}. The onset of the regime at which there is significant size dependence is largely coincident for both quantities when sweeping down the temperature. Given that the pairing correlations below $T_c$ have an algebraic decay, $C(r) \sim r^{-\eta(T)}$  with $\eta(0) = 0$ and $\eta(T_c)=0.25$, one obtains that the two-dimensional pair structure factor in a finite system of linear size $L$ scales as $P_s(L) = L^{2-\eta(T)}f(L/\xi_{\rm BKT})$, with $\xi_{\rm BKT} \propto \exp[b/\sqrt{|T-T_c|}]$~\cite{Moreo1991,Paiva2004}, where $b$ is a non-universal constant. 

Figure~\ref{fig:attractive}(g) gives this scaling using a temperature-adjusted  $\eta(T)$~\cite{Scalettar1989,SM}. The best parameters $b$ and $T_c$ are  extracted via the cost function ${\cal C}_{P_s}$, as before, and are shown in Fig.~\ref{fig:attractive}(i). Based on this, we similarly show a scaling analysis of the average sign of individual weights $\langle {\cal S}_\sigma\rangle$ to a KT form, see Fig.~\ref{fig:attractive}(h), with corresponding cost function ${\cal C}_{\langle {\cal S}_\sigma\rangle}$ displayed in Fig.~\ref{fig:attractive}(j). Both quantities scale remarkably precisely; the minor discrepancy in $T_c$ [$T_c^{P_s} = 0.15(2)$ and  $T_c^{\langle {\cal S}_\sigma\rangle} = 0.18(3)$] can be accounted by the relatively wide temperature-region in which ${\cal C}$ is small. For both quantities, we take the smallest $b$ given the constraint  of best collapse in a smooth curve. Lastly, we note that allowing for the possibility that the non-universal parameter $b$ takes different  values below and above the transition when lowering the temperature (requiring thus a multi-dimensional minimization procedure), may improve the convergence of the estimations of $T_c$~\cite{Aramthottil2021}

\section{Discussion and Outlook}\label{sec:discussion_and_outlook}
We have shown that the spin-resolved sign of auxiliary-field QMC simulations can be used as a {\it quantitative} marker of quantum critical behavior. The total sign also exhibits a similar role, as suggested by the Ionic Hubbard model results (see SM~\cite{SM}), but the former has the benefit of being useful when symmetries prevent the occurrence of an overall sign problem. Our work lays the foundation for similar investigations of other models, especially ones that give rise to a (spin-resolved) \textit{phase problem}. This can arise either from the presence of imaginary terms in the Hamiltonian, as in the Kane-Mele Hubbard model~\cite{Hohenadler2012,Zheng2011}, or from the particular decoupling scheme used. That is precisely the case of SU(2) symmetric Hubbard-Stratonovich transformations~\cite{meng10}, but investigations in  Appendix~\ref{app:su2_HS} show that the averaged spin-resolved phase similarly tracks the onset of the ordered regime when approaching the thermodynamic limit for the SU(2) honeycomb Hubbard model. 

Furthermore, other Hamiltonians, such as the \textit{spinless} fermion Hubbard model in either the honeycomb~\cite{Wang2014} or square-lattice with a $\pi$-flux, which in the Majorana basis evade the sign problem~\cite{ZiXiang2015,li16}, can be studied by examining the average sign of the Pfaffian of a single weight in that basis~\footnote{The original total weight given by a single determinant in the single-particle basis is converted to a product of weights related by complex conjugation when using the Majorana basis to describe the Hamiltonian, hence no sign (phase) problem.}, similar to what we have done here~\cite{Gotz2022}. In our results, the investigation of these three important models emphasizes that the sign of the determinants, interpreted as a minimal correlation function, is sufficient to assess critical properties, circumventing what is usually employed to determine scaling properties of physically motivated quantities.

While our investigation leads to the conclusion that the average (spin-resolved) sign displays scaling properties associated with critical behavior, it is less clear to understand \textit{why} this happens. The goal of the next subsection is to prove this. The remaining subsections tackle the explanation of criticality of the $\langle {\cal S}_\sigma\rangle$, the value we used of the dynamic critical exponent and lastly we follow with an outlook for future studies.

\subsection{Demonstration of non-analyticity of 
$\langle {\cal S} \rangle$}

We provide here a formal proof of the non-analyticity of $\langle {\cal S} \rangle$ which provides a rigorous theoretical framework for our numerical results.  Consider the re-writing of the partition function ${\cal Z}$ associated with a statistical mechanics problem with degrees of freedom $\{x\}$ and weight $W(\{x\})$, via sampling instead with a modified weight, $W^\prime(\{x\})$:
\begin{align}
{\cal Z} &= \int D\{x\} \, W(\{x\})
         = \int D\{x\} \, \frac{W(\{x\})} {W{^\prime}(\{x\})} \,
         W{^\prime}(\{x\})
\nonumber \\
         &= \frac{\int D\{x\} \, \frac{W(\{x\})} {W{^\prime}(\{x\})} \,
         W{^\prime}(\{x\})}
{ \int D\{x\} \, W{^\prime}(\{x\}) }
\,\,
{ \int D\{x\} \, W{^\prime}(\{x\}) }
\nonumber \\
 &= \Big\langle \frac{W(\{x\})} {W{^\prime}(\{x\})} \Big\rangle^{\prime}  \,\,
{\cal Z}^{\prime} \ \ .
\label{eq:proof1}
\end{align}
Here ${\cal Z}^{\prime} $ is the partition function associated with the weight $W^{\prime}$ and the prime on $ \Big\langle \frac{W(\{x\})} {W{^\prime}(\{x\})} \Big\rangle^{\prime} $ implies a weighting with $W^{\prime}$.

If there is a thermal or quantum phase transition occurring at a critical point associated with the original weight $W$, then from Eq.~(\ref{eq:proof1}) it is clear that
the associated non-analyticity in ${\cal Z}$ (and in the corresponding free energy density) implies that either 
${\cal Z}^{\prime}$ or  $ \Big\langle \frac{W(\{x\})} {W{^\prime}(\{x\})} \Big\rangle^{\prime} $ is non-analytic {\it at the same critical value}. Under the assumption that
${\cal Z}^{\prime}$ does not have the same critical point
(an unlikely coincidence) the non-analyticity must reside in
 $ \Big\langle \frac{W(\{x\})} {W{^\prime}(\{x\})} \Big\rangle^{\prime} $.
 
Let us now apply this general reasoning to the sign problem. There $W^{\prime}=|W|$ and $ \Big\langle \frac{W(\{x\})} {W{^\prime}(\{x\})} \Big\rangle^{\prime} $ is the average sign $\langle {\cal S} \rangle$. {\it Our conclusion is that a critical point in the underlying model implies critical behavior in this average sign.} We note that Eq.~(\ref{eq:proof1}) is nothing more than a re-writing of the well-known observation that the average sign is the exponential of the difference between the free energies ${\cal F}$ and ${\cal F}^{\prime}$ associated with the weights $W$ and $W^{\prime}$. However, this re-writing more clearly exposes the behavior of the average sign at a critical point.

Despite the simplicity of the argument, there are three important points to clarify. The first is that the particular value of the average sign is not universal. This is, of course, well-known. In the auxiliary field Quantum Monte Carlo method, $\langle {\cal S} \rangle$ depends on the particular Hubbard-Stratonovich transformation employed. What is universal, however, is that $\langle {\cal S} \rangle$ is non-analytic at the critical point of the model defined by $W$ (again, under the assumption of the absence of an `accidental' situation in which $W^{\prime}$ shares the precise same critical value)\footnote{Note that if one employs a different Hubbard-Stratonovich transformation, with say more degrees of freedom as in the case of the SU(2)-symmetric one explored in Appendix~\ref{app:su2_HS}, not only the weights are different but the integrand variable $\{x\}$ is also changed. This does not affect the above-stated conclusion, which is independent of the form of degrees of freedom being summed.}.

The second observation is that while a critical point implies a non-analyticity of $\langle {\cal S} \rangle$, the converse is not necessarily true. That is, a sign problem does not imply the existence of a critical point [see, e.g., Ref.~\cite{Schoof2015} for the uniform electron gas].  This also is known to be the case: the single-site Hubbard model has a sign problem with an anomalous Hubbard-Stratonovich transformation~\cite{batrouni90}, even though it manifestly has a completely well-behaved partition function. This does not reduce the potential utility of $\langle {\cal S} \rangle$ in locating a critical point. An analogy is useful. A single (Ising) spin in an external magnetic field $B$ has a non-zero magnetization $m$. But that a non-zero $m$ can occur in a trivial situation certainly does not imply that a (spontaneous) non-zero $m$ is uninformative concerning the occurrence of a magnetic phase transition. So too, here, the fact that $1-\langle {\cal S} \rangle $ can become non-zero in trivial situations does not make it unable to discern phase transitions.

The third remark concerns the non-analyticity of $\cal Z$, which is only observed in the thermodynamic limit: As for physical observables, the partition function is always analytic in finite systems~\footnote{Here we imply, of course, that a first-order phase transition is not under consideration, where non-analyticities do occur even within finite system sizes.}. For example, in the `textbook' problem of the magnetic phase transition of the two-dimensional Ising model, while large lattice sizes exhibit a peaked behavior of either the specific heat or the magnetic susceptibility close to the critical temperature below which order ensues, proper non-analytic behavior is only seen in approaching the thermodynamic limit, where such peaks approach
divergent behavior.  However, this does not prevent one from obtaining critical exponents by carefully scaling the results for the existing system sizes. The same rationale is valid mutatis mutandis to the partition function ${\cal Z}$: Only in the $N\to \infty$ limit does it show non-analyticity at the critical point. In models where one remaps the weights, as in the cases affected by the sign problem, it is then immediate to realize that while the non-analyticity is imprinted in $\langle {\cal S}\rangle$ in the thermodynamic limit, scaling of this quantity in finite-system sizes allows the extraction of the critical exponent, as we perform here.

In summary, the fact that the partition function (or the free energy) exhibits singular behavior thus implies that almost any observable will inherit the singularity as well. In the case of the sign, in particular, we have a formal proof of inheritance, as exposed in Eq.~\ref{eq:proof1}.

\subsection{Spin-resolved sign criticality}
A similar rationale can be derived in the case of the spin-resolved sign. For example, in a bipartite lattice at half-filling [the first model we investigate, the SU(2) honeycomb Hubbard model], it is then easy to show that weights associated with each fermionic flavor are related: $W_{\overline\sigma}(\{ x\}) = C_{\{ x\}}W_{\sigma}(\{ x\})$, where $C$ is a {${\{ x\}}$}-dependent positive constant ($=e^{\lambda\sum_{i\tau}x_{i\tau}}$, with ${\rm cosh} \, \lambda = e^{|U| \Delta \tau / 2}$)~\cite{Hirsch1985}\footnote{In the case of the attractive Hubbard model, the constant $C_{\{x\}}=1 \ \ \forall\{x\}$ provided the single-particle part of the Hamiltonian is the same for both spin species when using the charge decomposition Hubbard-Stratonovich transformation in the interaction terms.}. Therefore, the average sign of either of the weights reads
\begin{eqnarray}
     \langle {\cal S}_\sigma\rangle&=&\frac{\sum_{\{x\}} {\rm sgn}(W_\sigma(\{ x\})  ) C_{\{ x\}}[W_\sigma(\{ x\})]^2}{\sum_{\{x\}} C_{\{ x\}} [W_\sigma(\{ x\})]^2} \nonumber \\ 
     &\equiv&\frac{\sum_{\{x\}} {\rm sgn}(\sqrt{\rho(x)}) \rho(x)}{\sum_{\{x\}} \rho(x)} \equiv \frac{\sum_{\{x\}}  \rho^\prime(x)}{\sum_{\{x\}} \rho(x)}\nonumber \\
     &\equiv& \frac{{\cal Z}^\prime}{\cal Z}\ .
     \label{eq:spin_resolved_non_analyt}
\end{eqnarray}

Thus provided that a potential non-analyticity in the modified partition function ${\cal Z}^\prime$ does not coincide with the one for the original problem ${\cal Z}$, similarly to Eq.~\ref{eq:proof1}, this dictates that $\langle {\cal S}_\sigma\rangle$ should exhibit non-analytic behavior when ${\cal Z}$ does.
An interesting observation concerns the cases where a symmetry relates the spin-resolved signs in such a way
that the total sign remains at unity.  In that case, the non-analyticity {\it must} originate in the spin-resolution.
This emphasizes that even in `protected' cases, an analysis of the (spin-resolved) sign could still provide
insight into critical behavior.

\begin{figure}[htp!]
\centering
\includegraphics[width=1\columnwidth]{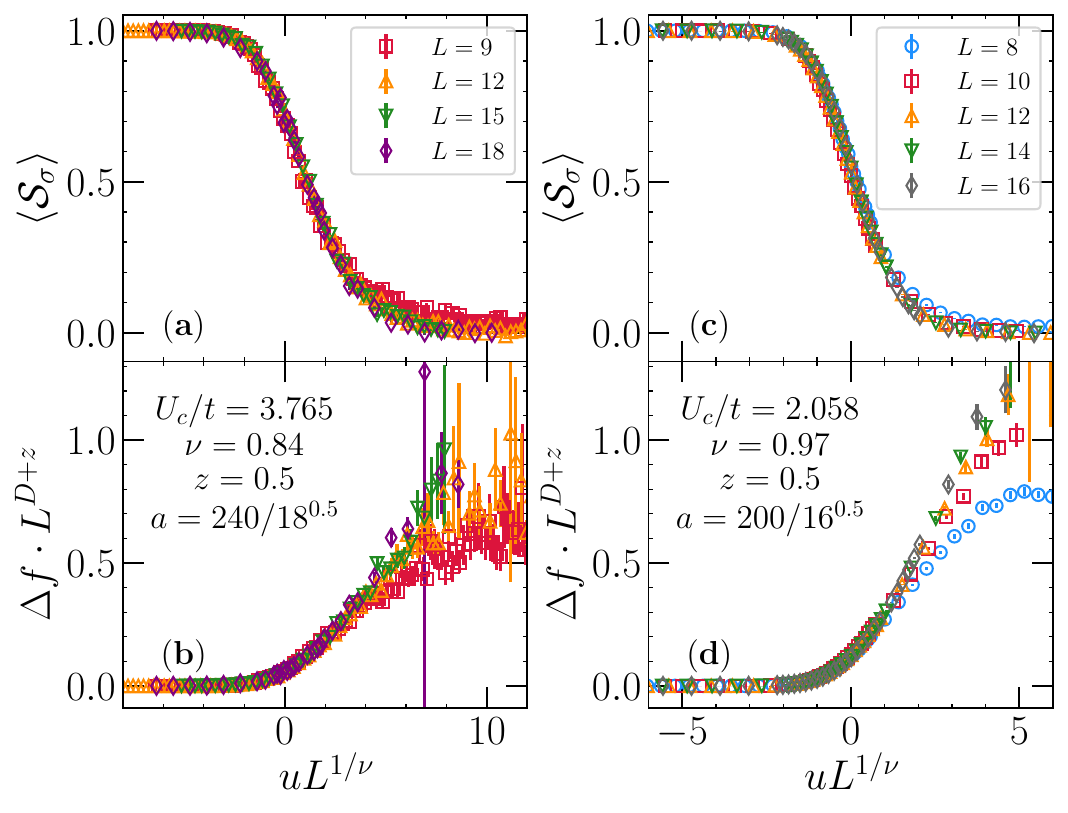}
\caption{Scaling analysis of the spin-resolved sign and the corresponding difference of free energies for the SU(2) honeycomb Hubbard model [(a) and (b)] and the square lattice SU(2) Ionic Hubbard model [(c) and (d)]. Parameters used are indicated, with imaginary-time discretization $t\Delta\tau = 0.1$.}
\label{fig:Delta_scaling}
\end{figure}

From Eq.~\ref{eq:spin_resolved_non_analyt} the logic follows the same as the one with a standard sign problem: One can define this ratio as the ratio of exponentials of corresponding free energy densities of probability distributions $\rho$ and $\rho^\prime$:
\begin{equation}
    \langle {\cal S}_\sigma\rangle = e^{-\beta V(f_{\rho^\prime} - f_\rho)}\ .
\end{equation}
For the case of the quantum phase transitions we have investigated, we demonstrated  \textit{numerically} that this quantity satisfies the scaling ansatz in the vicinity of the quantum critical point, i.e., $\langle {\cal S}_\sigma\rangle = g(uL^{1/\nu}, L_\tau/L^z)$. Consequently, the difference in free energy densities reads:
\begin{equation}
    \Delta f \equiv f_{\rho^\prime} - f_\rho= -\frac{1}{L_\tau\Delta\tau\cdot L^D}\log[ g(uL^{1/\nu},L_\tau/L^z)]\ .
\end{equation}
From Fig.~\ref{fig:scaling_hc}(d), we notice that $g(uL^{1/\nu}, L_\tau/L^z)$ goes from 1 to 0 when $u\simeq 0$, consequently $\Delta f $ shows a departure from zero at this same point. That is, the difference in free energy densities, initially zero in the non-interacting regime and in the weakly correlated one, turns finite when approaching the Mott phase as if the free energy densities of the models with probability distributions $\rho^\prime$ and $\rho$ undergo a `transition' to distinct values. Finally, as we fix the ratio $L_\tau/L^z =a$,
\begin{equation}
    \Delta f = -\frac{1}{a\Delta\tau\cdot L^{D+z}}\log[ g(uL^{1/\nu},a)].
\end{equation}
Figures \ref{fig:Delta_scaling}(a) and \ref{fig:Delta_scaling}(b)  summarize this reasoning for the SU(2) honeycomb Hubbard model, showing the scaling of the difference of free energy densities, where we emphasize that $\Delta F \equiv \Delta f \cdot L^{D+z}$ \textit{is} a function of the interaction strength~\footnote{From a historic perspective on the study of the sign problem in DQMC it is easy to see how this was missed: in both the one-dimensional and two-dimensional square lattice Hubbard models, the critical interaction is $U_c/t = 0^+$, hence at any finite interaction the difference in free energy densities is positive. The honeycomb Hubbard model (and perhaps others with $U_c > 0$) is unique in this aspect in that it allows one to precisely understand the regime where the free energy densities of the two systems differentiate (start to diverge), and its relation to the onset of the ordered phase.}. Similar logic applies to the Ionic Hubbard model [Figs.~\ref{fig:Delta_scaling}(c) and \ref{fig:Delta_scaling}(d)], in spite of the partial weights no longer being trivially related. That is, within the band-insulating regime, the difference in free energy $\Delta F$ of the two distributions is zero, deviating from each other once the correlated metal phase at $U = U_c$ is approached. This confirms the critical behavior we numerically observe for these models derives from the non-analyticity of the partition function in the critical point [Eq.~\eqref{eq:proof1}] that becomes imprinted in the average (spin-resolved) sign.

\subsection{Dynamic critical exponent}\label{sec:z}
One of the aspects of the scaling analysis of $\langle {\cal S}_\sigma\rangle$ that defies current expectations relates to the value of the dynamical critical exponent $z$ we have used. In particular, for the SU(2) honeycomb Hubbard model, field-theory predictions assert $z=1$~\cite{Herbut2009, Herbut2009b}, and numerical simulations using projective quantum Monte Carlo methods that directly tackle the $T=0$ limit often use this as a starting point~\cite{Sorella2012,Assaad2013,Otsuka2016}. Our simulations, on the other hand, employ the corresponding finite-temperature version of this algorithm~\cite{Blankenbecler1981,Hirsch1985}, such that the ground-state physics is only obtained asymptotically when $\beta\to\infty$ or when the typical correlation lengths $\xi$ are sufficiently large such that they are comparable to the linear system size $L$~\cite{Hirsch1985}. Verification of the latter is possible by examining the $\beta$-dependence of the antiferromagnetic structure factor
\begin{equation}
    S_{\rm AF} = \frac{1}{2L^2}\sum_{i,j} {(-1)}^{\delta}  \langle (\hat n_{i,\uparrow}-\hat n_{i,\downarrow}) (\hat n_{j,\uparrow}-\hat n_{j,\downarrow})\rangle
\end{equation}
with $\delta = 0$ ($\delta = 1$) if sites $i$ and $j$ are in the same (different) sublattice.

\begin{figure}[htp!]
\centering
\includegraphics[width=0.9\columnwidth]{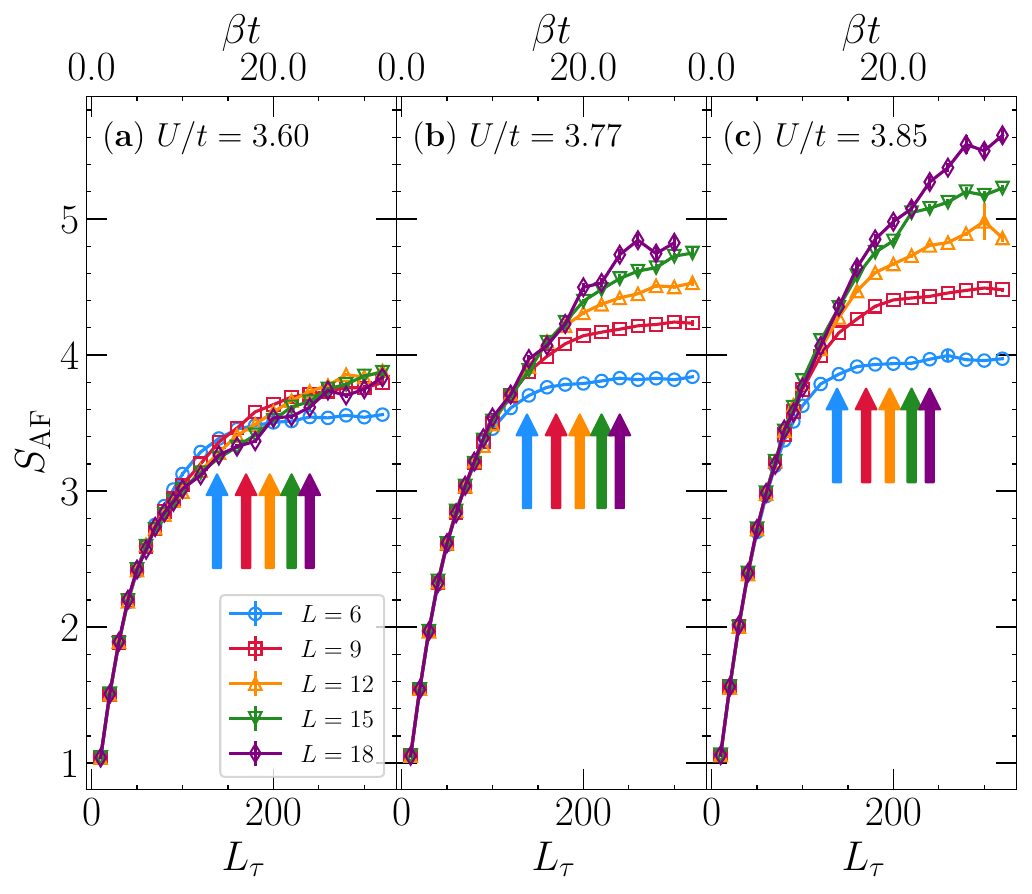}
\caption{Dependence of the antiferromagnetic structure factor on the inverse temperature $\beta$ (number of imaginary-time slices $L_\tau$, fixing $t\Delta\tau=0.1$) for the SU(2) honeycomb Hubbard model in the vicinity of the quantum critical point: (a) $U/t=3.60$, (b) $U/t=3.77$ and (c) $U/t=3.85$. The vertical arrows mark the value of $L_\tau$ used in the scaling analysis of $\langle {\cal S}_\sigma\rangle$, color-matching with the corresponding linear system size $L$.
}
\label{fig:S_AF_vs_Ltau}
\end{figure}

A saturation of $S_{\rm AF}$ with increasing $\beta$ indicates that $\xi \simeq L$~\cite{paiva05}, and is readily obtained deep in the ordered phase. Close to $U_c$, however, the observation of such a saturation demands (numerically) prohibitive values of $L_\tau$, as indicated in Fig.~\ref{fig:S_AF_vs_Ltau}. As a result, the currently employed values of the imaginary-time slices in the scaling analysis of $\langle {\cal S}_\sigma\rangle$ [marked by the arrows in Fig.~\ref{fig:S_AF_vs_Ltau}] inevitably lead to the conclusion that finite-temperature effects are at play here, and the observed scaling relates to a low-but-finite temperature crossover that emanates from the quantum phase transition. Consequently, the dynamical critical exponent need not be pinned at $z=1$, and the value we use, obtained after scaling of $L_\tau/L^z$ for the current range of imaginary-time slices employed,
endows the ability to study the quantum criticality. In other words, the fact that we adjust the dynamical critical exponent for the current range of temperatures is what allows one to obtain numerically accurate values of the pair $(U_c,\nu)$.

While this may come as a surprise, it becomes more clear after performing a scaling of a physical quantity, in particular the one which dictates the onset of magnetic ordering at the quantum critical point, the antiferromagnetic structure factor, as shown in Appendix~\ref{app:S_AF}. There one finds that a $z<1$ (in practice $z \simeq 0.5$) gives the best data collapse, and that the same combination of $(U_c,\nu)$ which scales the spin-resolved sign is seen to similarly scale $S_{\rm AF}$.

\subsection{Outlook}
In summary, our determination of the quantum criticality via the scaling of the sign in many models is within the range of existing investigations. Because the literature shows considerable variation in the quantitative location of the different QCP's~\cite{Toldin2015,Otsuka2016,paris07,bouadim07,Moreo1991,Paiva2004}, the results here offer an alternative, and potentially more accurate, route of resolving a challenging problem in correlated electron models. A similar investigation can be carried out with methods that directly tackle the ground-state limit, and we defer this analysis to a future study. Turning to the thermal transitions, a final, more speculative, line of inquiry is to investigate the potential existence of a Kosterlitz-Thouless transition in the \textit{repulsive} Hubbard model (and its variants) away from half-filling via the analysis of the average sign when entering the $\langle {\cal S}\rangle\to0$ phase in this Hamiltonian~\cite{mondaini2021}. A preliminary study is given in the SM~\cite{SM}, affirmatively indicating this connection. 
 
\begin{acknowledgments}
\noindent We acknowledge discussions with I.~Herbut, S.~Sachdev, Z.Y.~Meng, and R.~Singh. R.M.~acknowledges support from the National Natural Science Foundation of China (NSFC) Grants No.~U2230402, 12111530010, 12222401, and No.~11974039. R.T.S.~was supported by the grant DE‐SC0014671 funded by the U.S. Department of Energy, Office of Science. Computations were performed on the Tianhe-2JK at the Beijing Computational Science Research Center.
\end{acknowledgments}

\vskip0.02in
\appendix

\section{Methods}
\label{app:Methods}
\vskip0.02in
In all calculations, we make use of the finite-temperature determinant quantum Monte Carlo method~\cite{Blankenbecler1981,Hirsch1985}. Via a sequence of Trotterization, Hubbard-Stratonovich decoupling of quartic terms by means of the introduction  of an auxiliary field~\cite{Hirsch1983}, and final fermionic integration, the partition function is written in terms of the determinants of $N\times N$ matrices $M_\sigma$ [where $N$ is the number of sites in Eq.~\eqref{eq:ham_spinful}] for each spin-component $\sigma$. These are the weights $W( \{ x\} )$ referred to in the main text. Instead of summing over all configurations of the field $\{ x\} \to \{ x_{i\tau}\}$, importance sampling is performed while observing the statistical convergence of both physical observables (when possible) and the average sign of the weights. The \textit{only} approximation used is the imaginary-time discretization $\Delta\tau$ which we take as $1/10$ for the quantum transitions or $1/16$ for the thermal ones. Statistical sampling varies among the different models, but in all cases an average of the results are taken for each individual set of parameters over dozens of independent samplings (typically from 20 to 48), with thousands of Monte Carlo sweeps for each run.

In order to decouple the interactions in all SU(2) models we investigate, we apply either the spin-decomposed Hubbard-Stratonovich transformation~\cite{Hirsch1983,Loh1992},
\begin{align}\label{eq:HS_sz}
e^{-\Delta \tau U (\hat n_{i\uparrow} - \frac{1}{2}) 
(\hat n_{i\downarrow} - \frac{1}{2})} 
= \frac{1}{2} e^{-U \Delta \tau/4} \sum_{x_i=\pm 1} e^{\lambda x_i (\hat n_{i\uparrow} - \hat n_{i\downarrow})}
\, , 
\end{align}
for repulsive interactions ($U>0$), or its counterpart (charge decomposition)
\begin{align}
e^{-\Delta \tau U (\hat n_{i\uparrow} - \frac{1}{2}) 
(\hat n_{i\downarrow} - \frac{1}{2})} 
= \frac{1}{2} e^{-|U| \Delta \tau/4} \sum_{x_i=\pm 1} e^{\lambda x_i (\hat n_{i\uparrow} + \hat n_{i\downarrow} - 1)}
\, , 
\end{align}
in the case that $U<0$. In both transformations, ${\rm cosh} \, \lambda = e^{|U| \Delta \tau / 2}$. Finally, the matrices $M_\sigma$ entering in the weights read
\begin{align}
M_\sigma = \mathds{1} + B_{\sigma,L_\tau}B_{\sigma,L_\tau-1}\ldots B_{\sigma,1},
\label{eq:M_matrix}
\end{align}
with $B_{\sigma,\tau} = e^{K}e^{V_{\sigma,\tau}}$. Here, $K$ is an imaginary-time independent $N\times N$ matrix containing all one-body terms in the Hamiltonian (including hopping and chemical potential), whose entries are multiplied by $-\Delta\tau$. In turn, $V_{\sigma,\tau}$ is diagonal with entries that depend on the Hubbard-Stratonovich transformation used. For the repulsive case, $V^{ii}_{\sigma,\tau} = \lambda\sigma x_{i\tau}$ ($\sigma =\pm 1$ for $\uparrow$ and $\downarrow$), while $V^{ii}_{\uparrow,\tau} = V^{ii}_{\downarrow,\tau} = \lambda x_{i\tau}$ for attractive interactions.

\vskip0.02in
\section{Eigenvalues of the $M_\sigma$ matrices}
\label{app:Matrix}
\vskip0.02in

A possibility to infer numerically that the signs of the determinants can track phase transitions is via the analysis of the spectrum of the $M_\sigma$ matrices, as defined in Eq.~\ref{eq:M_matrix}, whose determinant gives the partial weight of a certain configuration $\{x\}$. Similarly, one can define this matrix in its space-time formulation~\cite{Blankenbecler1981},
\begin{equation}
    M_\sigma(\{x\}) = \begin{bmatrix}
                    \mathds{1} &  &  &  & B_{\sigma,{L_\tau}} \\
                    -B_{\sigma,1} & \mathds{1} &  &  &  \\
                     & -B_{\sigma,2} & \mathds{1} &  &  \\
                     &  & \ddots & \ddots &  \\
                     &  &  & -B_{\sigma,{L_\tau-1}} & \mathds{1} 
                    \end{bmatrix}, \nonumber
\label{eq:M_space_time}
\end{equation}
by structuring the $N\times N$ matrices $B_{\sigma,\tau}$, the single-particle propagators, as defined in the Appendix~\ref{app:Methods}. A drawback is that $M_\sigma(\{x\})$ has now dimensions $(NL_\tau)\times(NL_\tau)$, but one of the benefits of this representation is that the range of eigenvalues is now shrunken while preserving the value of the determinant. Besides that, it is numerically stable since no matrix multiplications among $B_{\sigma}$'s are necessary to build it.

\begin{figure*}[t!]
\centering
\includegraphics[width=0.72\textwidth]{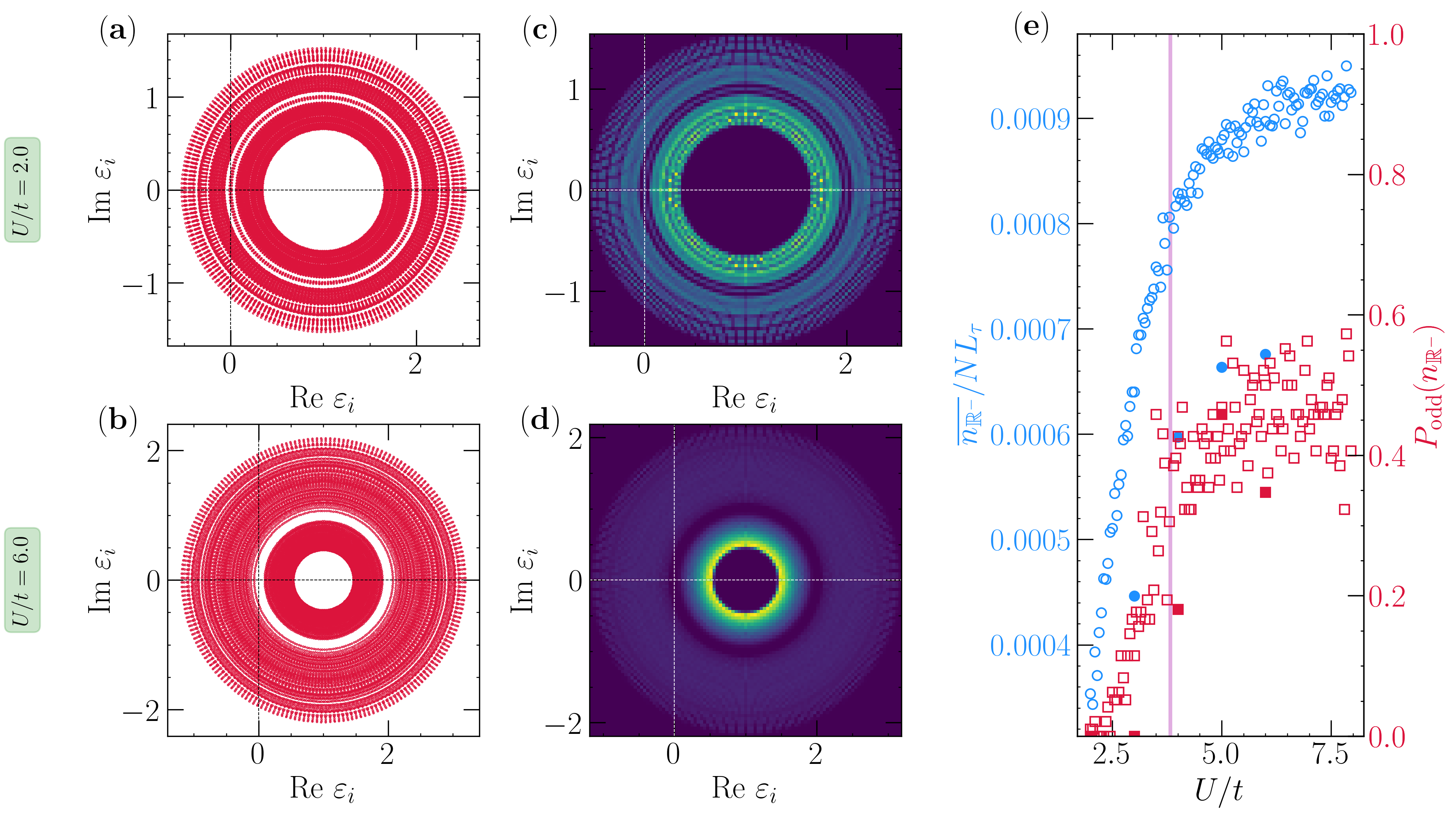}
\caption{The eigenspectrum of the $M_\sigma(\{x\})$ matrix represented in the complex plane for a single Hubbard-Stratonovich configuration $\{x_{i,\tau}\}$ extracted in the Monte Carlo sampling, for $U/t=2$ (a) and $U/t=6$ (b). (c) and (d) exhibit a two-dimensional histogram of the eigenspectrum when combining results of 96 configurations -- brighter colors display a larger counting. In (a--d), the linear lattice size is $L=9$ and $L_\tau=200$, such that each configuration leads to $32\ 400$ eigenvalues. (e) shows both the (normalized) number of eigenvalues which are in the negative real axis (these dictate whether there is a spin-resolved sign problem or not) and the fraction of the configurations that possess an odd number of eigenvalues in $\mathbb{R}^-$. Empty (full) symbols refer to $L=6$ ($L=9$); the vertical shaded region gives the confidence interval of the quantum critical point location. As in the main text, here we use $t\, \Delta\tau=0.1$.
}
\label{fig:eigenv_M}
\end{figure*}

Focusing on the SU(2) honeycomb Hubbard model, we start by analyzing in Fig.~\ref{fig:eigenv_M} the spectrum $\{\varepsilon_i\}$ of $M_\sigma(\{x\})$ for values of the interactions far below ($U/t=2$) and far above ($U/t=6$) the transition point $U_c$. Figures \ref{fig:eigenv_M} (a) and \ref{fig:eigenv_M} (b) show that a structural transition occurs in the eigenvalue spectrum, here computed for a single typical configuration of the auxiliary field $\{x_{i,\tau}\}$. This observed structural transition is generic, as shown by the corresponding two-dimensional histograms in Figs.~\ref{fig:eigenv_M} (c) and (d), obtained by combining eigenvalues of four field configurations `visited' over the course of the Monte Carlo sampling for 24 independently seeded Markov chains, resulting in 96 configurations in total.

While the quantum phase transition is hinted in the eigenvalues of $M_\sigma(\{x\})$, so far we have not drawn a connection to the spin-resolved sign problem. Being a real matrix (for this model with the Hubbard Stratonovich transformation highlighted in the Appendix \ref{app:Methods}), its eigenvalues come as either complex conjugate pairs or real numbers. Since the determinant (product of eigenvalues) does not change sign when multiplying the conjugate pairs, a sign problem is only a function of the number of eigenvalues in the negative real axis, $n_{\mathbb{R}^-}$. That is, if $n_{\mathbb{R}^-}$ is odd (even), $\det M_\sigma(\{x\}) < 0$ ($>0$). Typically, $n_{\mathbb{R}^-}$ is very small in comparison to the total number of eigenvalues $2L^2 L_\tau$ in this Hamiltonian. Yet, it is a clear function of the interaction strength, growing at large $U/t$, as shown in Fig.~\ref{fig:eigenv_M}(e) by $\overline{n_{\mathbb{R}^-}}$, after averaging over different configurations $\{x\}$. Finally, the percentage of those configurations that possess an \textit{odd} number of eigenvalues in the negative real axis, $P_{\rm odd}(n_{\mathbb{R}^-})$, also grows reaching around 50$\%$ within the ordered phase. As a result, the average spin-resolved sign $\langle {\cal S}_\sigma\rangle$ converges to zero. 

It is currently unclear to us if a physical meaning can be attributed to the number of negative eigenvalues of $M_\sigma(\{x\})$, in terms of the fields $\{x_{i,\tau}\}$, and the winding of world lines they are associated with.

\vskip0.02in
\section{The scaling of $S_{\rm AF}$} 
\label{app:S_AF}
\begin{figure*}[t!]
\centering
\includegraphics[width=0.75\textwidth]{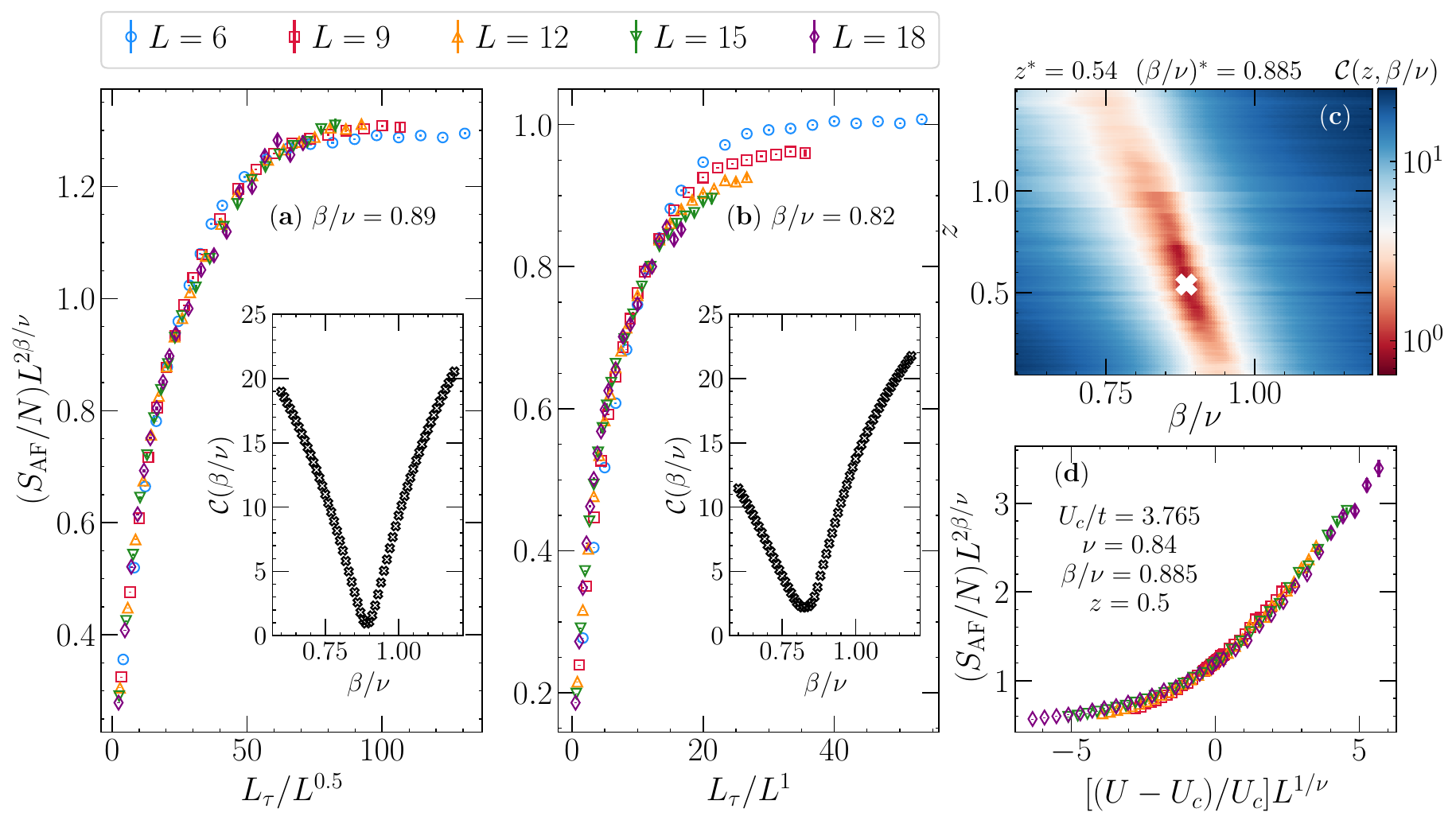}
\caption{The scaling analysis of the antiferromagnetic structure factor. In (a) and (b) the scaled $S_{\rm AF}$ dependence on $L_\tau/L^z$, with $z=0.5$ and 1, respectively. Insets give the corresponding cost function dependence on the ratio of exponents $\beta/\nu$. (c) Color-mesh plot of the cost function ${\cal C}(z, \beta/\nu)$, where the white marker at $(z^*, \beta/\nu^*) = (0.54, 0.885)$ depicts its minimum value. (d) The scaling collapse of $S_{\rm AF}$ with $uL^{1/\nu}$, with parameters extracted from the analysis of the spin-resolved sign in the main text [Fig.~\ref{fig:scaling_hc}]. The imaginary-time discretization used is $t \Delta\tau = 0.1$.
}
\label{fig:scaling_S_AF}
\end{figure*}

In the main text, we argue that owing to finite-temperature effects one needs to adjust the dynamical critical exponent $z$ from its expected $z=1$ value in order to perform a scaling analysis of the spin-resolved sign $\langle {\cal S}_\sigma \rangle$. While the scaling analysis we perform for $\langle {\cal S}_\sigma \rangle$ is quantitatively precise despite its novelty, similar constraints should apply to the case of the scaling of physical quantities.
Following this logic, we perform the scaling of the antiferromagnetic order parameter, $m_{\rm AF} = \lim_{L\to \infty}\sqrt{S_{\rm AF}/N}$, with $N=2L^2$ the number of sites of the SU(2) honeycomb Hubbard model, Eq.~\eqref{eq:ham_spinful}. This quantity follows a scaling ansatz of the form $m_{\rm AF} = L^{-\beta/\nu}g(uL^{1/\nu},L_\tau/L^z)$~\cite{Assaad2013,Otsuka2016}, which in turn implies the antiferromagnetic structure factor scales as
\begin{equation}
    \frac{S_{\rm AF}}{N} = L^{-2\beta/\nu}g(uL^{1/\nu},L_\tau/L^z)
    \label{eq:scaling_S_AF}
\end{equation}
Unlike previous studies that tackle this same transition using $T=0$ quantum Monte Carlo methods~\cite{Sorella2012,Assaad2013,Toldin2015,Otsuka2016}, we make the $L_\tau/L^z$ dependence explicit in order to account for a finite-$T$ influence on the scaling.

We start by fixing $U/t = 3.77$, such that the dependence on the first argument of the scaling function in Eq.~\eqref{eq:scaling_S_AF} is negligible, i.e., $u \simeq0$ \footnote{Notice that this value is chosen based on the critical value $U_c/t = 3.765$ that comes from the scaling of the spin-resolved sign.}. Figure~\ref{fig:scaling_S_AF}(a) and \ref{fig:scaling_S_AF}(b) display the scaled structure factor vs.~$L_\tau/L^z$ by fixing $z=0.5$ and $1.0$, respectively, while adjusting the ratio of exponents $\beta/\nu$ that gives the best data collapse. The latter is obtained by the minimization of the cost function ${\cal C}(\beta/\nu)$, whose definition is the same as given in the main text, and is displayed as insets in Figs.~\ref{fig:scaling_S_AF}(a) and \ref{fig:scaling_S_AF}(b). Notably, the data collapse is significantly better at $z\simeq0.5$ compared to the one for $z=1$, a first indication that a dynamic critical exponent smaller than one results in an improved scaling. Compiling the cost function in a range of $(z,\beta/\nu)$-values, shown in Fig.~\ref{fig:scaling_S_AF}(c) as a color-mesh plot, we obtain the minimum cost function at $z^*=0.54$ and $(\beta/\nu)^* = 0.885$. The latter is compatible with the value $\beta/\nu\simeq0.9$ obtained after the first-order $\epsilon$-expansion of the Gross-Neveu model~\cite{Herbut2009b, Assaad2013}. Lastly, by fixing this ratio of exponents $(\beta/\nu)$ while choosing the set of parameters used to perform the scaling of $\langle {\cal S}_\sigma\rangle$ in the main text, $(U_c/t, \nu, z) = (3.765, 0.84, 0.5)$, we report in Fig.~\ref{fig:scaling_S_AF}(e) the dependence of the scaled structure factor with respect to the first argument of the scaling function: The exhibited collapse is remarkably good, thus confirming that the average (spin-resolved) sign can indeed be used to infer criticality. As far as we are aware, this is the first time that a finite-temperature quantum Monte Carlo method was used to obtain critical exponents of a transition pertaining to the Gross-Neveu universality class, and the key step for its success is the tuning of the dynamic critical exponent $z$.

\vskip0.02in
\section{Other Hubbard-Stratonovich transformations}
\label{app:su2_HS}
\vskip0.02in
Our main results indicate that the (spin-resolved) average sign carries fundamental information about phase transitions and their universality classes. However, given that the sign problem is basis-dependent, that is, by choosing another Hubbard-Stratonovich transformation, the average sign of the quantum Monte Carlo weights can change~\cite{Hirsch1986,batrouni90}, an immediate question that arises is: can one still infer quantum critical points using ${\rm sgn}(W_\sigma(\{x\}))$? To answer it, we report further numerical tests.

\begin{figure}[t!]
\centering
\includegraphics[width=0.95\columnwidth]{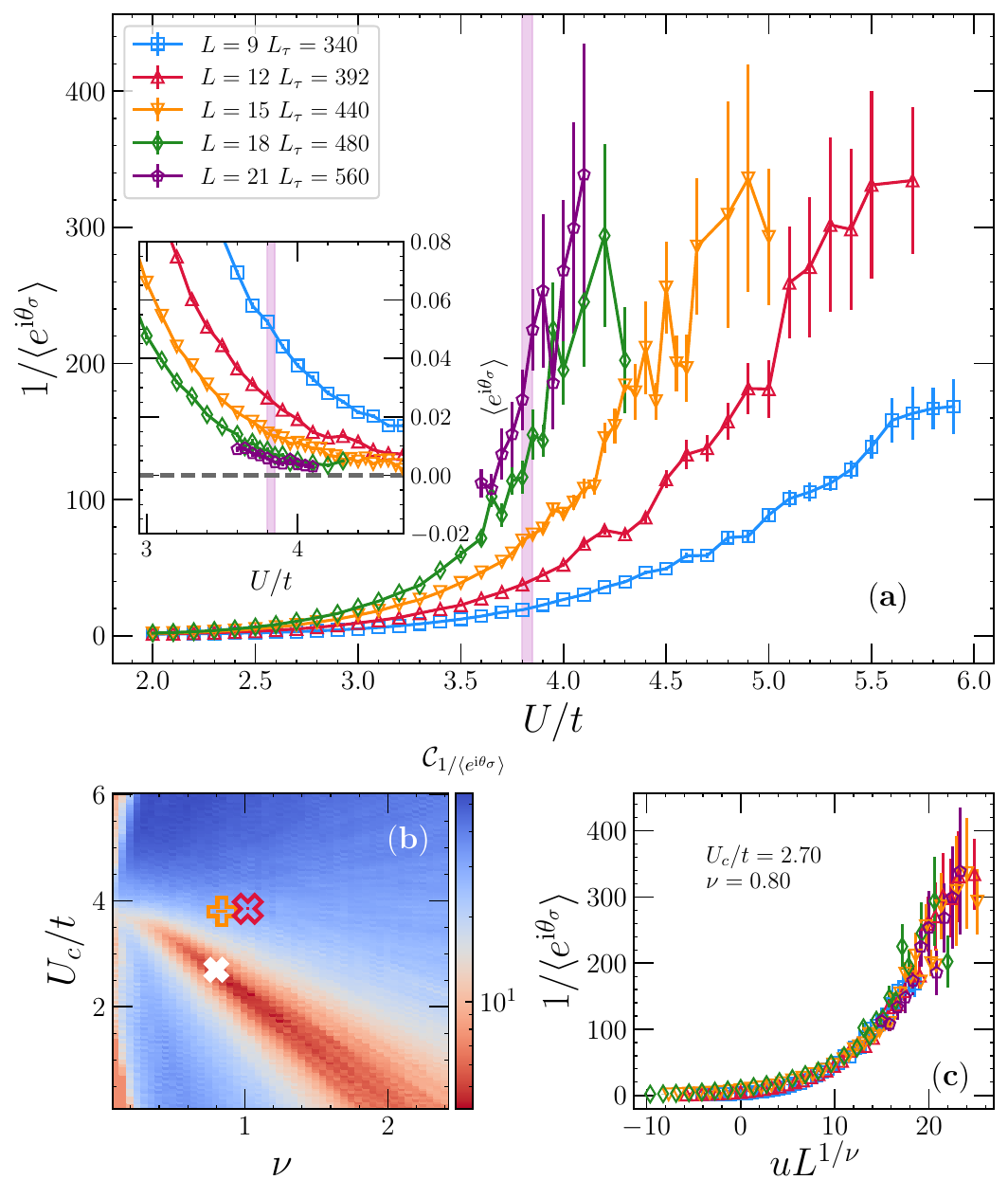}
\caption{(a) Dependence of the \tcc{inverse} average spin resolved phase on the interaction magnitude $U/t$ for the SU(2) honeycomb Hubbard model. We fix the ratio $L_\tau/L^z$ approximately constant, with $z=0.5$. \tcc{The inset is a zoom-in of the critical region while using $\langle e^{{\rm i}\theta_\sigma}\rangle$ instead; the vertical shaded region is the recent estimation in the literature of the QCP location~\cite{Toldin2015, Otsuka2016}. (b) gives the cost-function results for the scaling analysis of $1/\langle e^{{\rm i}\theta_\sigma}\rangle$, whereas (c) the best scaling collapse, leading to $(U_c/t, \nu) = (2.70, 0.80)$. The imaginary-time discretization used is $t \Delta\tau = 0.05$.}
}
\label{fig:ave_phase_vs_U}
\end{figure}

An often used Hubbard-Stratonovich transformation is one that explicitly conserves the SU(2) symmetry for each configuration $\{x_{i\tau}\}$ of the field~\cite{meng10,Zheng2011,Hohenadler2012}:
\begin{align}
&e^{-\Delta \tau U (\hat n_{i\uparrow} + \hat n_{i\downarrow} - 1)^2/2} = \nonumber \\ &\sum_{x_{i\tau}=\pm1,\pm2} \gamma(x_{i\tau}) \prod_{\sigma} e^{{\rm i} \sqrt{\Delta\tau U/2}\eta(x_{i\tau})(\hat n_{i\sigma} -1/2)}+{\cal O}(\Delta\tau^4)\ .
\label{eq:su2_hs}
\end{align}
It comes at the expense of having a four-valued discrete field $x_{i\tau}=\pm1,\pm2$, accompanied by a few (real) constants,
\begin{align}
    \gamma(\pm1)=1+\sqrt{6}/3\ ;\  
    \eta(\pm1)=\pm\sqrt{2(3-\sqrt{6})} \nonumber\\
    \gamma(\pm2)=1-\sqrt{6}/3\ ; \ 
    \eta(\pm2)=\pm\sqrt{2(3+\sqrt{6})}\ .
\end{align}
In such case, the transformation is not strictly exact but brings an error proportional to ${\cal O}(\Delta\tau^4)$, negligible in practice in comparison to the one introduced by the Trotter decomposition of the one and two-body terms in the Hamiltonian, ${\cal O}(\Delta\tau^2)$. 

Given its form, the Monte Carlo weights can become complex, in principle, but in the SU(2) honeycomb Hubbard model, owing to its bipartite nature, one can show that the weights associated with the two-spin components are complex conjugate pairs at half-filling~\cite{meng10, Zheng2011, Hohenadler2012}. As a result, no \textit{phase problem} emerges. Nonetheless, this does not guarantee that the average phase of each fermionic flavor $\langle e^{{\rm i}\theta_\sigma}\rangle$ needs to be \textit{real} and raises the immediate question of whether the sign still captures information about the onset of an ordered phase. Numerical simulations we performed, however, point out affirmatively to both: ${\rm Im} \langle e^ {{\rm i}\theta_\sigma}\rangle \rightarrow 0$ throughout the sampling, and that $\langle e^ {{\rm i}\theta_\sigma}\rangle (U=U_c)\rightarrow 0$ when approaching the thermodynamic limit. The latter is reported in Fig.~\ref{fig:ave_phase_vs_U}(a) (see inset), using  $L_\tau/L^{1/2} = \frac{240}{18^{1/2}}$ approximately fixed for different system sizes.

Unlike in the case of a spin-decomposed Hubbard-Stratonovich transformation \tcc{[Eq.~\eqref{eq:HS_sz}]}, where a crossing of $\langle {\cal S}_\sigma\rangle$ for different system sizes leads to immediate identification of $U_c$, here the nature of the dependence of the average spin-resolved phase with $U$ makes a scaling process more challenging. \tcc{While}, the trend of $\langle e^{{\rm i}\theta_\sigma}\rangle$ with different system sizes suggests that the average phase converges towards zero when approaching the quantum critical point \tcc{(or that $1/\langle e^{{\rm i}\theta_\sigma}\rangle$ diverges at $U\to U_c$), a scaling analysis similar to that performed in the main text results in a significant underestimation of the critical interaction $U_c$ [Fig.~\ref{fig:ave_phase_vs_U}(b) and \ref{fig:ave_phase_vs_U}(c)]; the critical exponent $\nu$, on the other hand, is close to most recent predictions~\cite{Toldin2015, Otsuka2016}. We note that our original arguments regarding the non-analytic behavior of the spin-resolved sign should carry over to the spin-resolved phase. That is, considering that the total weight is decomposed in $W(x)=W_\sigma(x)W_{\overline\sigma}(x)$}
\tcc{\begin{eqnarray}
\langle e^{{\rm i}\theta_\sigma}\rangle_{|W|} &=& \frac{\sum_x e^{{\rm i}\theta_\sigma(x)}|W_\sigma(x)W_{\overline\sigma}(x)|}{\sum_x |W_\sigma(x)W_{\overline\sigma}(x)|} \times \frac{{\cal Z}_W}{{\cal Z}_W} \nonumber \\
&=& \langle e^{{\rm i}\theta}\rangle_{|W|} \frac{\sum_x e^{{\rm i}\theta_\sigma(x)}|W_\sigma(x)W_{\overline\sigma}(x)|}{{\cal Z}_W}\ ;
\end{eqnarray}
and that $W_{\overline\sigma}(x)=W_\sigma^*(x)$ (i.e., $\langle e^{{\rm i}\theta}\rangle_{|W|} = 1$), the partition function of the original model reads}
\tcc{\begin{equation}
    {\cal Z}_W = \frac{1}{\langle e^{{\rm i}\theta_\sigma}\rangle_{|W|}}\cdot {\cal Z^\prime} \ \ \text{where}\ \ {\cal Z}^\prime \equiv \sum_x e^{{\rm i}\theta_\sigma(x)}|W_\sigma(x)|^2\ .
\end{equation}
As a result, non-analytic behavior in the thermodynamic limit that appears in ${\cal Z}_W$ at the critical point is guaranteed to be reflected in the averaged spin-resolved phase provided the modified partition function ${\cal Z}^\prime$ is sufficiently analytic in the vicinity of $U_c$.}

Future investigations with both larger sizes and improved statistics may settle the possible determination of critical exponents in this case. \tcc{Yet, as will become clear in the following Appendix (Appendix~\ref{app:dtau_analysis}), an explanation for this mismatch of the values of $U_c$ possibly stems from the fact that the average spin-resolved phases $\langle e^{{\rm i}\theta_\sigma}\rangle$ suffer from substantially larger dependence on the value of the imaginary-time discretization, in comparison to the spin-resolved sign $\langle {\cal S}_\sigma\rangle$ studied in the main text.}

\tcc{\section{Dependence on the imaginary-time discretization $\Delta\tau$} \label{app:dtau_analysis}}
\tcc{Other than statistical accuracy, which can always be systematically improved, our quantum Monte Carlo simulations suffer from only one bias: the discrete imaginary-time $\Delta\tau$. It derives from the single approximation employed in the method when using a Trotter decomposition to isolate the quartic terms of the Hamiltonian in writing the partition function~\cite{Hirsch1985, dossantos03}. As previously established, in doing so, one ends up with errors proportional to ${\cal O}(\Delta\tau^2)$. While it is common to verify the discretization errors on physical quantities, noting how they converge in the limit $\Delta\tau\to0$ to establish the critical properties~\cite{meng10, Sorella2012, Otsuka2016, Assaad2013}, much less scrutiny is put on the dependence of the average sign/phases of the weights.}

\begin{figure}[t!]
 \vspace{0.6cm}
\centering
\includegraphics[width=0.99\columnwidth]{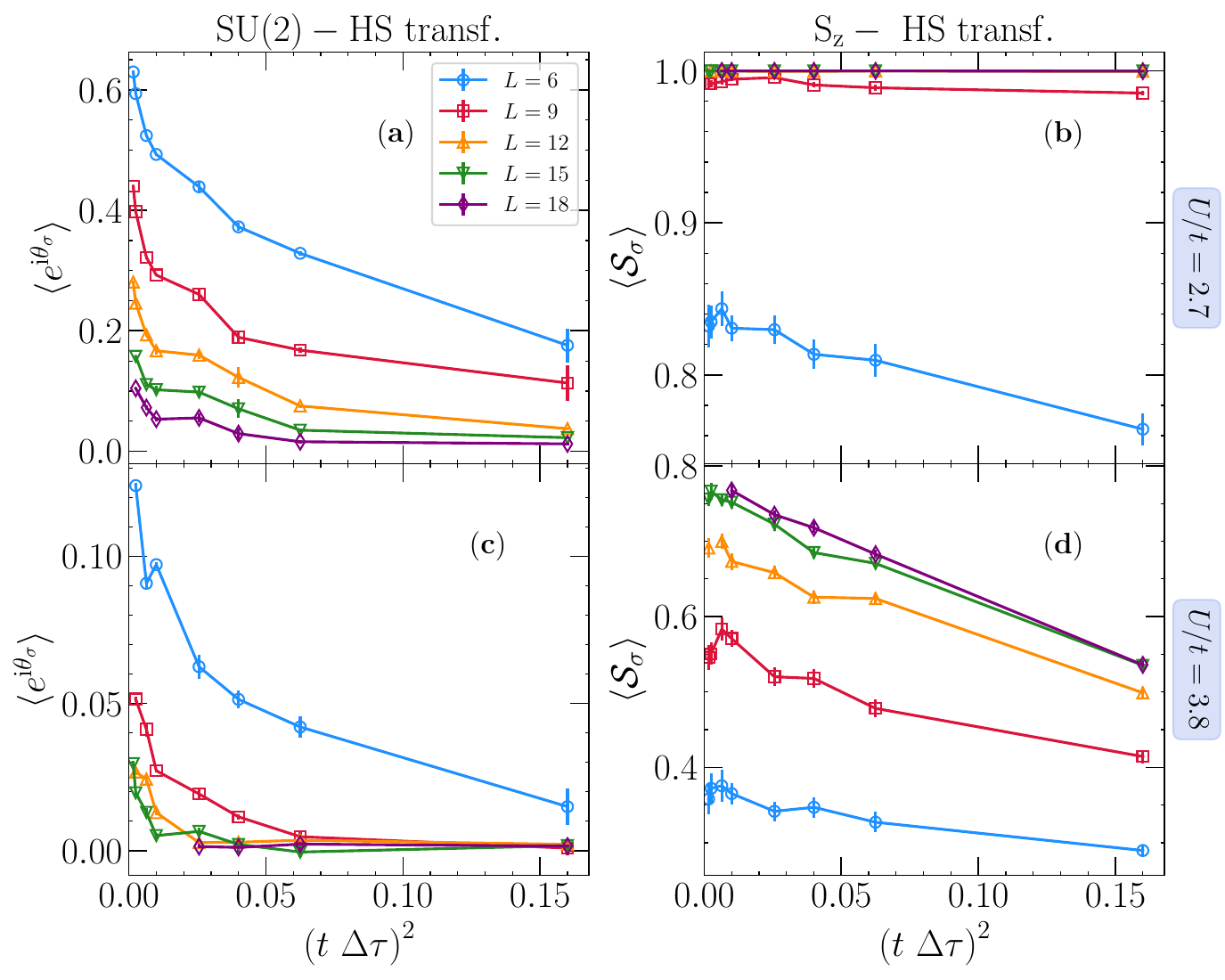}
\caption{\tcc{Dependence of the average spin-resolved phase $\langle e^{{\rm i}\theta_\sigma}\rangle$ [(a), (c)] and average spin-resolved sign $\langle {\cal S}_\sigma\rangle$ [(b), (d)] on the square of the imaginary-time discretization for the SU(2) honeycomb Hubbard model. We compare two values of the interaction strength, $U/t = 2.7$ [(a), (b)] and $3.8$ [(c), (d)], and contrast two types of Hubbard-Stratonovich transformations [Eqs.~\eqref{eq:HS_sz} and \eqref{eq:su2_hs}]. Close to saturation is observed for $\langle {\cal S}_\sigma\rangle$ in approaching $\Delta\tau\to0$, where statistical errors encompass the small variation observed. Instead, a substantial dependence is seen in this limit for $\langle e^{{\rm i}\theta_\sigma}\rangle$. All data are extracted at fixed inverse temperatures $\beta t = 20$.
}}
\label{fig:dtau_dependence}
\end{figure}

\tcc{To fill this gap, we report in Fig.~\ref{fig:dtau_dependence} the dependence of $\langle e^{{\rm i}\theta_\sigma}\rangle$ and $\langle {\cal S}_\sigma\rangle$ for the SU(2) honeycomb Hubbard model, using two values of the interactions $U/t = 2.7$ and 3.8. While the spin-resolved sign closely follows the linear dependence with $(t\Delta\tau)^2$, the same cannot be said about the spin-resolved phase. Here, $\langle e^{{\rm i}\theta_\sigma}\rangle$ has a substantial variation on the discretization used in the limit $\Delta\tau\to0$, which significantly compromises an estimation of the critical properties via the scaling analysis we propose. This prevents us from obtaining accurate values of $U_c$ and $\nu$ for the Dirac semi-metal to antiferromagnetic Mott insulator transition in Appendix~\ref{app:su2_HS} for this model. While the corresponding Hubbard-Stratonovich transformation [Eq.~\eqref{app:su2_HS}] introduces an extra error proportional to ${\cal O}(\Delta\tau^4)$, this clearly cannot explain the large dependence observed. It is currently unclear why such behavior emerges [physical quantities display the usual linear dependence in $(t\Delta\tau^2)$ at small $\Delta\tau$], and further investigations with a broader class of transformations are likely required to understand it. We leave this for future studies.}

\bibliography{signscaling}

\clearpage

\renewcommand{\theequation}{S\arabic{equation}}
\setcounter{equation}{0}

\onecolumngrid

\begin{center}

{\large \bf Supplementary Materials:
 \\Universality and Critical Exponents of the Fermion Sign Problem}\\

\vspace{0.3cm}

\end{center}

\vspace{0.6cm}

\beginsupplement

In these Supplementary Materials we provide additional data to support the scalings used, and try different scaling procedures as well. In addition, we show the scaling of the total sign $\langle {\cal S}\rangle$ for the Ionic Hubbard model to argue that this quantity follows $\langle {\cal S}_\sigma\rangle$  as long as it is not protected by some symmetry of the problem. Lastly, further details of the attractive Hubbard model are presented, discussing the problematic finite-size effects that occur in relatively small lattices and the temperature-dependence of the correlation exponent $\eta$. We end by providing preliminary data suggestive of a KT-scaling form on both $\langle {\cal S}_\sigma\rangle$ and $\langle {\cal S}\rangle$ for the repulsive Hubbard model at densities below half-filling.

\tableofcontents

\section{Scaling of the total sign in the Ionic Hubbard model}
\begin{figure}[h!]
\centering
\includegraphics[width=0.59\columnwidth]{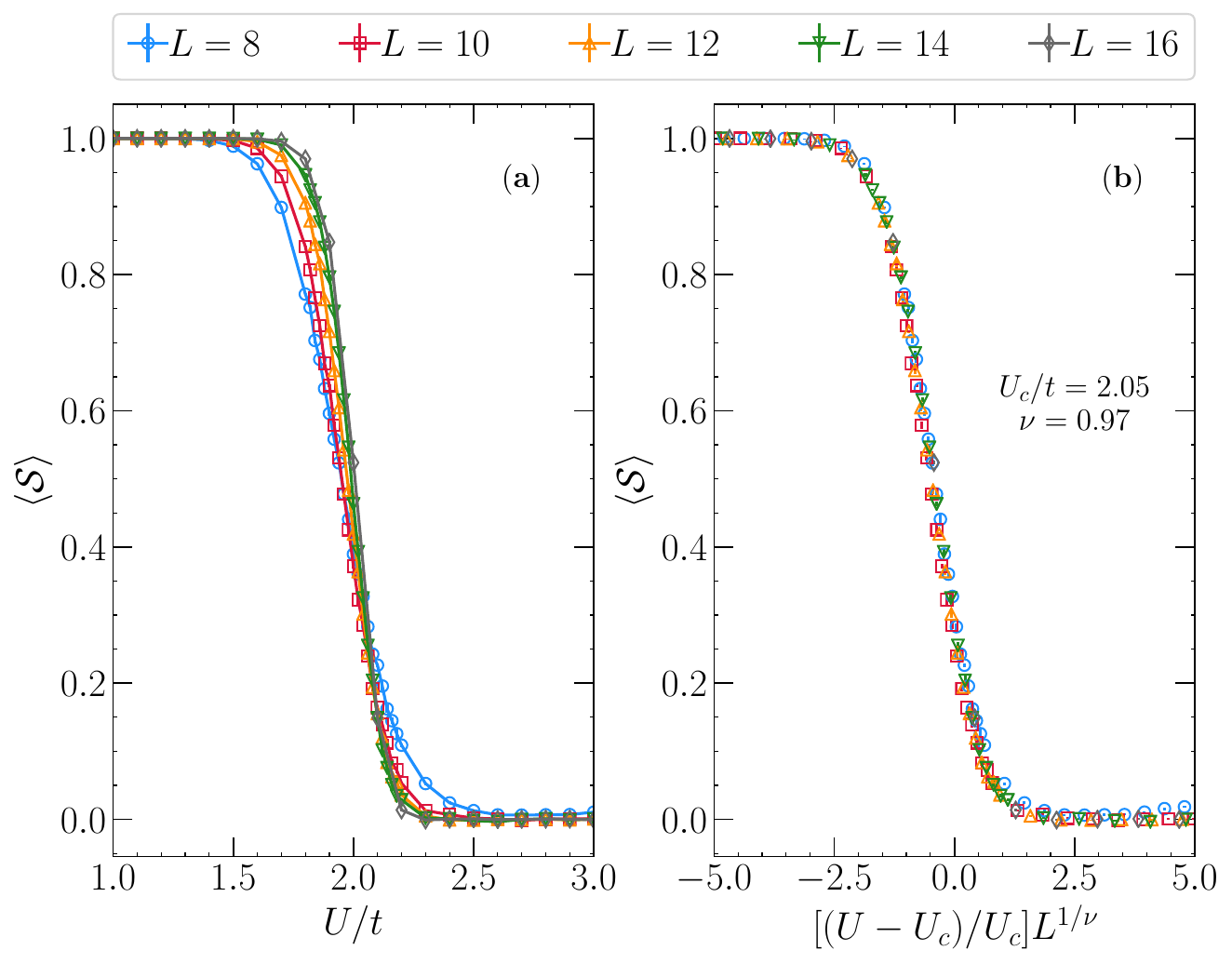}
\caption{Scaling analysis of the \textit{total} sign $\langle {\cal S}\rangle$ of the QMC weights in the SU(2) Ionic Hubbard model. We use the same parameters obtained for the scaling of $\langle {\cal S}_\sigma\rangle$ in the main text, $U_c/t = 2.05$ and $\nu = 0.97$. As before, the $L_\tau/L^{0.5}$ ratio is fixed at 50, and $\Delta/t = 0.5$.}
\label{fig:total_sign_ionic_N_vs_U_scaling}
\end{figure}
Since we focus attention on the scaling of the sign of an \textit{individual} determinant in the main text, here we show that the same scaling properties are valid in the case of the \textit{total} sign of the QMC weights. This is possible in the case of the Ionic Hubbard model precisely because there is no symmetry-protection that guarantees the positive-definiteness of the total sign. We report in Fig.~\ref{fig:total_sign_ionic_N_vs_U_scaling} (a) the dependence of the average sign of the product of determinants (total weight of quantum configurations in QMC calculations) as a function of the interaction's strength. Similar to the case of spin-resolved sign, $\langle {\cal S}\rangle\to 0$ in the ordered phase, displaying a crossing at the critical interaction $U_c/t$ when considering different system sizes. That the critical interactions and correlation exponent $\nu$ extracted from the scaling of $\langle {\cal S}_\sigma\rangle$ similarly scales the curves for the total sign demonstrates the generality of the interpretation that the sign of the weights already hosts information about the critical properties of quantum models.

\section{Using a different scaling procedure}
\begin{figure}[ht!]
\centering
\includegraphics[width=0.7\columnwidth]{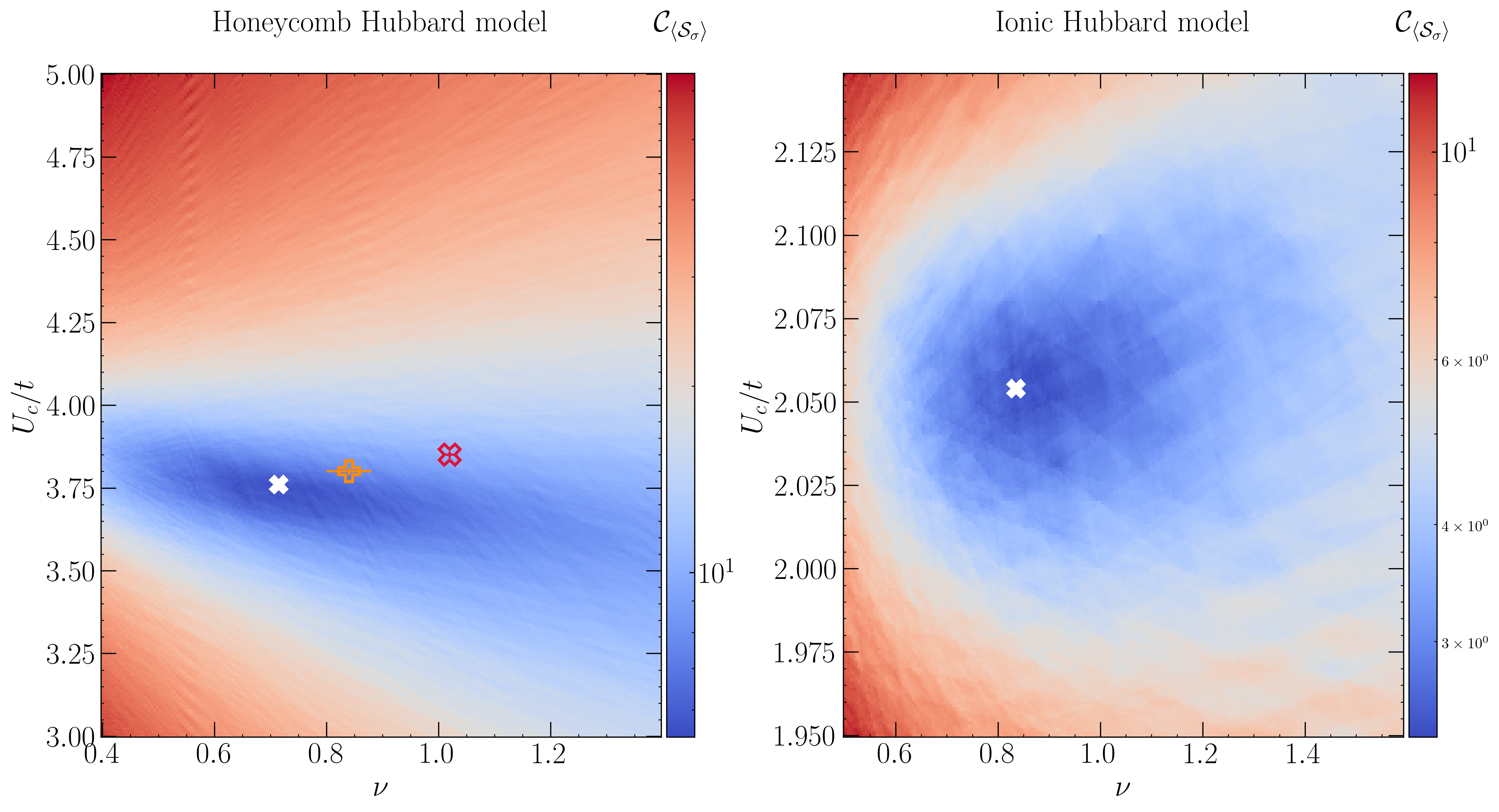}
\caption{Different scaling procedure. Cost function for the scaling of the spin resolved sign in the vicinity of the quantum phase transition for the Honeycomb Hubbard model (left) and the Ionic Hubbard model on the square lattice (right). The critical set of parameters $(\nu,U_c)$ is given by the white marker and corresponds to the point where ${\cal C}_{\langle {\cal S}_\sigma\rangle}$ is minimized. Cross markers in the left panel, as in Fig.~\ref{fig:scaling_hc}(c), display the results of Refs.~\onlinecite{Toldin2015,Otsuka2016}. The results are $(\nu,U_c/t) = (0.71, 3.73)$ for the honeycomb case, and $(\nu,U_c/t) = (0.83, 2.06)$ in the ionic model.}
\label{fig:cost_func_hc_ionic}
\end{figure}

In computing the critical exponent $\nu$ and the critical interactions $U_c$ that characterize and drive the Dirac semi-metal to Mott insulator transition, we used an analysis that relies on the fitting quality with a high-order polynomial in the vicinity of the transition. Here, we demonstrate that such results are hardly changed if we use different approaches. In particular, we try the cost-function method used in the extraction of the dynamic critical exponents $z$, and of the critical temperature in the KT-transition for the attractive Hubbard model.  The benefits of this procedure arise from the fact that `flat' regions in the data set do not contribute to the goodness of the fit, unlike in a polynomial approach, which may be challenging when dealing with both flat and decaying regimes of the data, which often incurs in problems of over-fitting it.

Figure~\ref{fig:cost_func_hc_ionic} displays the cost function ${\cal C}_{\langle {\cal S}_\sigma\rangle} = \sum_j (|y_{j+1} - y_j|)/(\max\{y_j\} - \min\{y_j\})-1$, where $y_j$ are the values of $\langle {\cal S}_\sigma\rangle(L_\tau,L)$, ordered according to their $[(U-U_c)/U_c]L^{1/\nu}$ value~\cite{Suntajs2020}. This is equivalent to Fig.~\ref{fig:scaling_hc}(e) and \ref{fig:ionic}(e), but using a different collapse method. In both cases the critical exponents are fairly close to their previous predictions (see caption of Fig.~\ref{fig:cost_func_hc_ionic}). Small discrepancies highlight the sensitivity of scaling analyses to the data treatment method. Yet, for the honeycomb Hubbard model where accurate predictions using physical observables are known~\cite{Toldin2015,Otsuka2016}, a different scaling method produces similarly good results.

\section{More details on the attractive Hubbard model}
\begin{figure}[ht!]
\centering
\includegraphics[width=0.7\columnwidth]{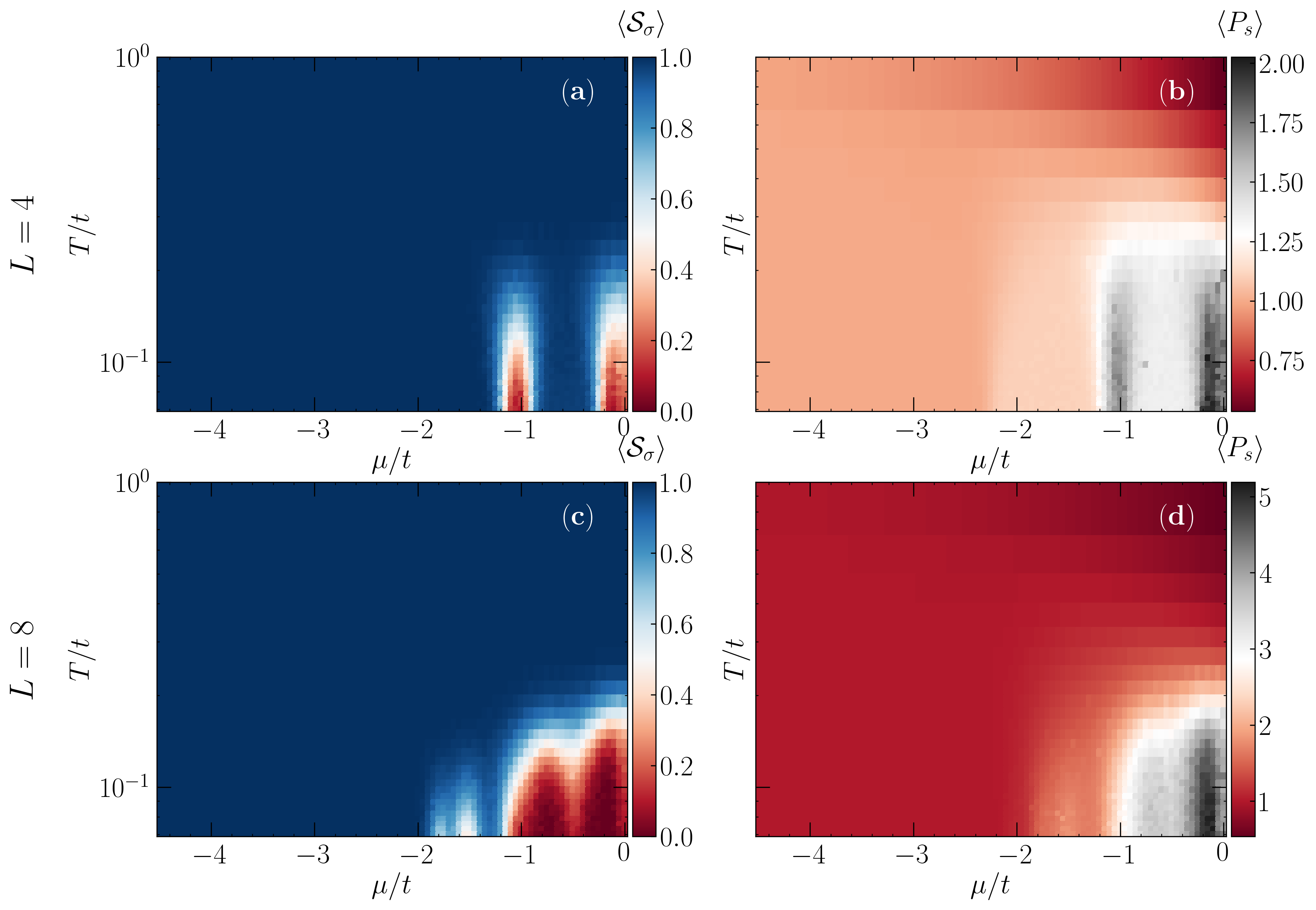}
\caption{Severe finite-size effects on the SU(2) attractive Hubbard model. Spin resolved sign and pair structure factor in the Hubbard model for small lattice sizes $L=4$ (a, b) and 8 (c, d) in the space of parameters $T/t$ vs. $\mu/t$. The dome structure in which $\langle {\cal S}_\sigma\rangle\to0$ and $P_s$ is large tracks the regimes where a $-\vec k, \vec k$ `nesting' of the Fermi surfaces occurs. Here interaction strength is $U/t = -4$.}
\label{fig:small_size_attractive}
\end{figure}

\paragraph{Severe finite-size effects at small lattice sizes.---} The attractive Hubbard model (analyzed in detail in the main text concerning the relationship between the spin-resolved average sign and superconducting behavior) has the peculiarity of being especially prone to the manifestation of finite-size effects in comparison to the other models investigated. In Fig.~\ref{fig:small_size_attractive}, we show the dependence of $\langle {\cal S}_\sigma\rangle$ and $P_s$ in the $T/t$ vs. $\mu/t$ plane for smaller lattice sizes than the ones considered for the KT scaling analysis. It is clear now that only a few special values of $\mu/t$ lead to an enhancement of the $s$-wave pairing at sufficiently low temperatures.  Nevertheless,
the structures are also reproduced by the onset of regions where $\langle {\cal S}_\sigma\rangle\to0$. One can understand this non-monotonic behavior by recalling that in small lattices, the coarse grained discretization of the Brillouin zone implies that a few special $\mu/t$ lead to a perfect nesting condition, where the Fermi surface crosses one of the allowed momentum points. Remarkably this occurs even at the intermediate values of $U/t = 4$ as seen in Fig.~\ref{fig:small_size_attractive}. Such oscillations have been noted before~\cite{Moreo1991}, and using small lattice sizes to extract the critical temperature can be responsible for large deviations from the expected behavior in the thermodynamic limit as shown in Ref.~\onlinecite{Paiva2004}.

\begin{figure}[ht!]
\centering
\includegraphics[width=0.7\columnwidth]{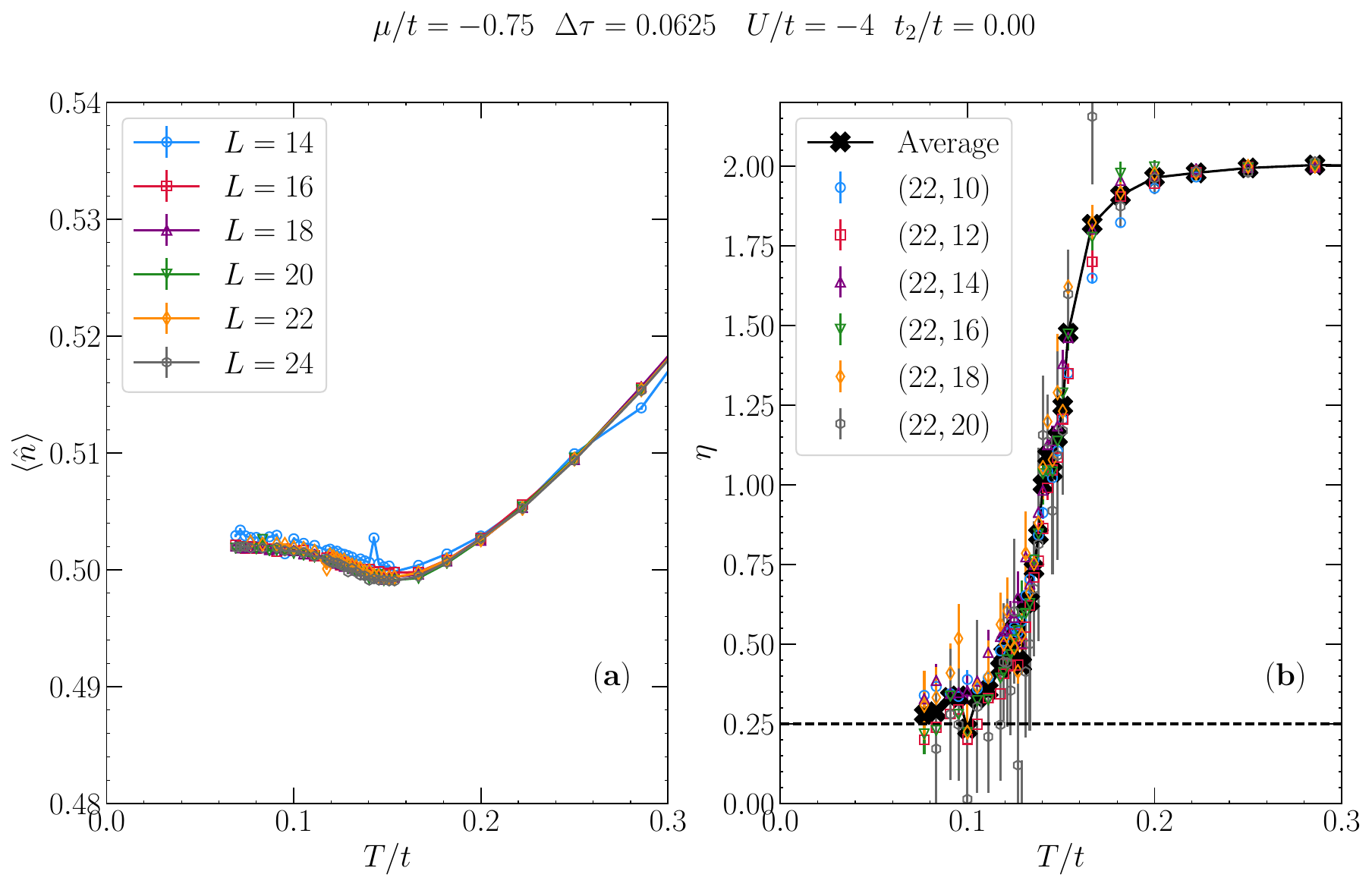}
\caption{Extra results of the SU(2) attractive Hubbard model. The resulting density as the temperature is decreased (a) and the estimation of the temperature-dependent exponent $\eta$ (b) which governs the real-space decay of the pairing correlations (see text). As in previous analyses, interaction strength is $U/t = -4$ and chemical potential is set at  $\mu/t = -0.75$.}
\label{fig:dens_eta}
\end{figure}

\paragraph{Estimating $\eta(T)$.---} In the main text, we perform a scaling analysis of the $s$-wave pair structure factor $P_s$ using the fact that in a finite system of linear size $L$ it should scale as $P_s(L) = L^{2-\eta(T)} f(L/\xi_{\rm KT})$~\cite{Moreo1991,Paiva2004}. For a  KT-type scaling, the exponent $\eta$ is temperature-dependent, and related to the decay of the real-space pair correlations $C(r = |{\bf i}-{\bf j}|)\equiv\langle \Delta_{\bf i}^\dagger \Delta_{\bf j}^{ \phantom \dagger} + \Delta_{\bf j}^\dagger \Delta_{\bf i}^{ \phantom \dagger}\rangle \sim r^{-\eta(T)}$~\cite{Paiva2004}. At $T=0$, true long-range order is manifest and $\eta(0) = 0$. As one approaches $T_c$ from below, the exponent smoothly evolves until $\eta(T_c) = 0.25$. In analogy to what was performed in Ref.~\onlinecite{Scalettar1989} for the \textit{classical} two-dimensional Heisenberg model, we here extract $\eta(T)$ by dividing the scaling form above for different system sizes. If comparing lattices with linear sizes $L$ and $L^\prime$, the exponent is thus
\begin{equation}
\eta(T) = 2 - \frac{\ln[P_s(L,T)/P_s(L^\prime,T)]}{\ln(L/L^\prime)}.
\label{eq:eta_T}
\end{equation}
Thence, selecting a reference (large) lattice $L$, one can extract $\eta(T)$ by checking the ratio in Eq.~\ref{eq:eta_T} at each temperature. This procedure is illustrated in Fig.~\ref{fig:dens_eta}(b), where the reference lattice has $L=22$. While even at the relatively low temperatures we investigate it is unclear the extrapolation to $\eta(T\to 0) = 0$ is attainable, the second known limit, $\eta(T\to T_c) = 0.25$, is readily observed. At high temperatures, $\eta$ saturates at 2, which is consistent with the fact that the pair structure factor no longer has size-dependence in this regime [see Fig.~\ref{fig:attractive}(g)].

\paragraph{Fixed chemical potential.---} The properties of the Hubbard model are often studied in terms of a fixed density of particles. When performing a grand-canonical, finite-temperature simulation, this is accomplished by setting the chemical potential in the regime of parameters that returns the desired electronic density. In particular, such an approach was applied before in studying the attractive Hubbard model away from half-filling~\cite{Scalettar1989,Moreo1991,Paiva2004}. Here, we  take a different route and fix the chemical potential (in the main text, $\mu/t=-0.75$). As displayed in Fig.~\ref{fig:dens_eta}(a), this procedure results in a slight deviation ($\sim 1\%$) of the densities around quarter-filling ($\langle \hat n\rangle = 0.5$) in the salient region where the KT transition takes place, with minimal size effects. As the critical temperature $T_c$ possesses minor variations around this density range~\cite{Paiva2004}, such a procedure is justifiable.

\section{The repulsive Hubbard model away from half-filling}
The implication that an estimation of the superconducting critical temperature $T_c$ can be extracted by the regime where the average sign of single weights in the Monte Carlo sampling drops to zero (furthermore obeying a KT-scaling form), immediately opens room for the exploration of the same procedure to tackle the repulsive case, which, unlike the attractive Hubbard model, displays a sign problem for the total weights (thus preventing using physical observables to track an eventual finite-temperature transition~\cite{mondaini2021}). We here use refinements of the standard repulsive Hubbard model [Eq.~\eqref{eq:ham_spinful}], by including a next-nearest neighbor hopping $t_2$. This has been shown to provide more accurate comparisons to experiments on cuprates~\cite{Piazza2012,Hirayama2018,Hirayama2019}.

\begin{figure}[ht!]
\centering
\includegraphics[width=0.7\columnwidth]{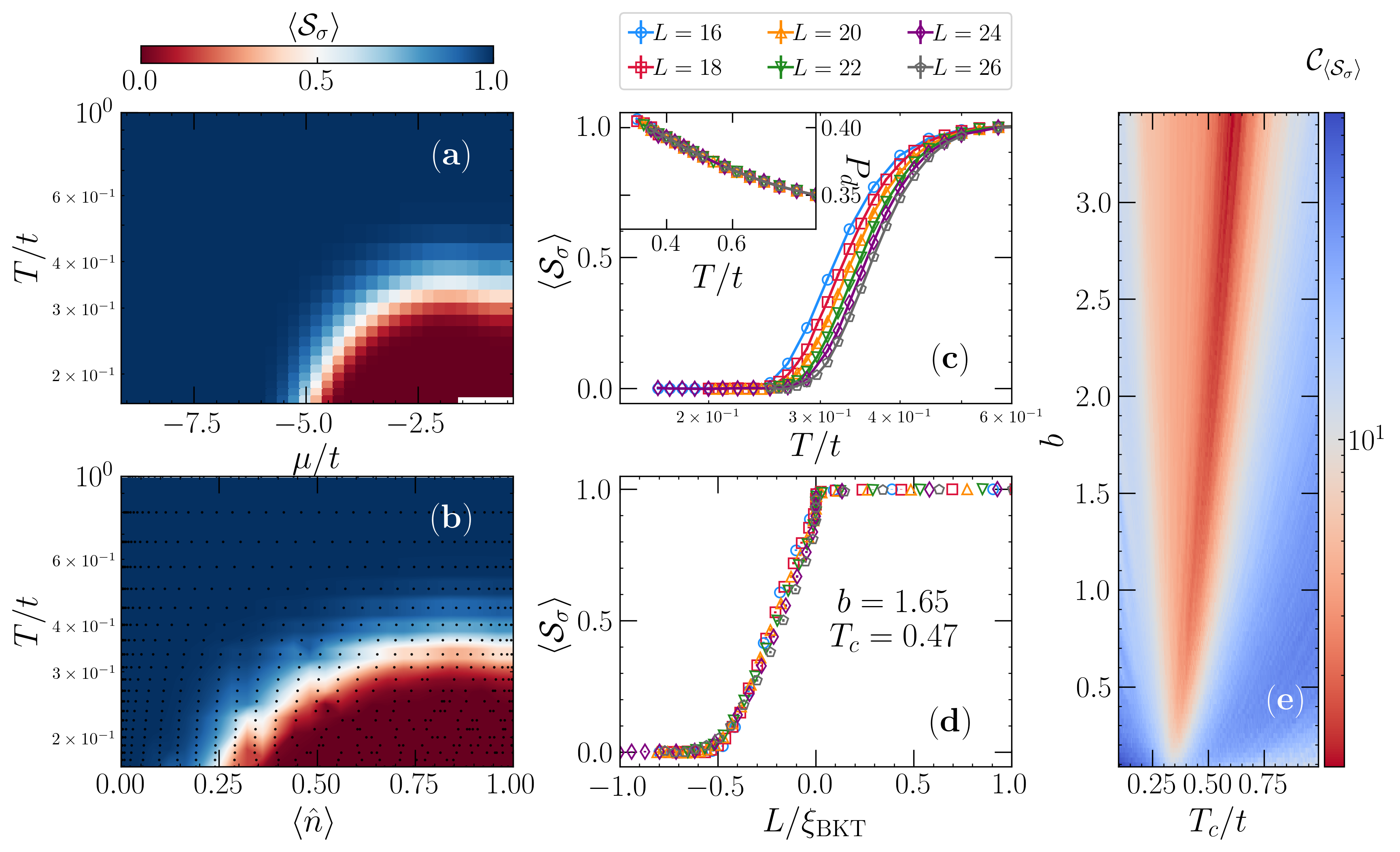}
\caption{The scaling of the spin-resolved sign in the repulsive Hubbard model. (a) The map of the spin resolved sign in the (a) $T$ vs.~$\mu$ and in the (b) $T$ vs.~$\langle \hat n \rangle$ planes. In the latter, black markers depict the densities extracted from the regular $\mu$ grid, and where the interpolation of the data is performed. (c) A cut of $\langle {\cal S}_\sigma\rangle$ at chemical potential $\mu/t = -2$, corresponding to density $\sim 0.8$. (d) A KT-scaling form ($\xi_{\rm BKT} = \exp\{b/\sqrt{|T-T_c|}\}$) with parameters as displayed, using lattice sizes ranging from $L=16$ to 26. (e) The corresponding cost function of the spin-resolved sign in the regime of parameters $(T_c,b)$. The inset in (c) displays the $T$-dependence of the $d$-wave pair structure factor in the regime of parameters where $\langle {\cal S}\rangle >0.25$: system-size dependence is negligible at this regime, and shows that the extensive regime is not attainable up until $\langle {\cal S}_\sigma\rangle \to 0$. The parameters are: interaction strength $U/t = 6$; next-nearest neighbor hopping $t_2/t = -0.2$, and imaginary-time discretization is $\Delta\tau = 1/16$.}
\label{fig:repulsive}
\end{figure}

\begin{figure}[hb!]
\centering
\includegraphics[width=0.7\columnwidth]{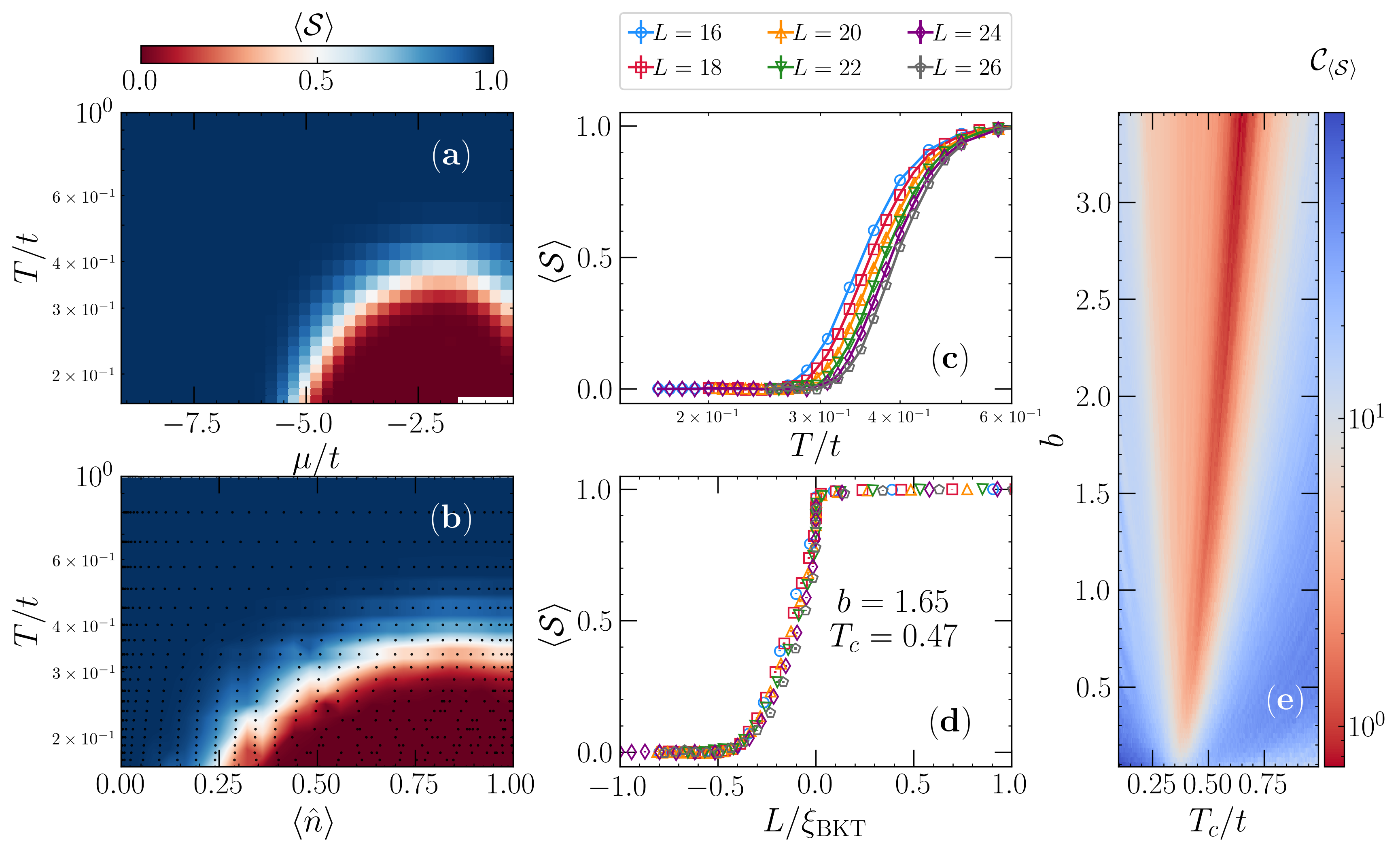}
\caption{The scaling of the total average sign in the repulsive Hubbard model. The same as in Fig.~\ref{fig:repulsive} but for the total sign $\langle {\cal S}\rangle$ instead of the spin-resolved one $\langle {\cal S}_\sigma\rangle$. The same set of parameters $b$ and $T_c$ scale both quantities equally well.}
\label{fig:repulsive_tot_sign}
\end{figure}

Figure~\ref{fig:repulsive} plots the `phase diagram' of the spin resolved sign [as in Fig.~\ref{fig:attractive}(a) and (b) in the main text for the attractive case] with varying temperatures and both chemical potentials [Fig.~\ref{fig:repulsive}(a)] and resulting electronic densities [Fig.~\ref{fig:repulsive}(b)]. The appearance of a $\langle {\cal S}_\sigma\rangle$-dome suggests the study of criticality is possibly analogous to the attractive case. By selecting a fixed chemical potential $\mu/t=-2$ (rendering densities $\langle \hat n \rangle \sim 0.8$ across a wide span of temperatures), we perform a KT-scaling analysis in Fig.~\ref{fig:repulsive}(c) and \ref{fig:repulsive}(d); the corresponding cost function is given in Fig.~\ref{fig:repulsive}(e).
The trend is remarkably similar to the $U<0$ model, where a finite critical temperature is known to occur. Indetermination in the location of $T_c$ arises from the fact that there is a wide regime in which $C_{\langle{ \cal S}\rangle_\sigma}(T_c,b)$ is small. Future studies on larger lattices, and considering the possibility that the non-universal parameter $b$ assumes different values above and below the putative transition, may render a more accurate estimation.

The repulsive Hubbard model at densities below unity allows us to further reinforce the argument that the total sign of the weights in the QMC calculations already provides information about critical behavior. As there is no symmetry protection that guarantees it to be pinned at 1, we can see in Fig.~\ref{fig:repulsive_tot_sign} that the same set of parameters that scale $\langle {\cal S}_\sigma\rangle$, equally well scale the total sign. Accompanied by similar results of the Ionic Hubbard model [see Fig.~\ref{fig:total_sign_ionic_N_vs_U_scaling}], we are led to the conclusion that in generic models without symmetry protection, the total sign reflects the same critical properties of the average sign of partial weights, either in quantum or thermal phase transitions.

Although the critical temperature obtained via the KT-scaling is remarkably higher than one would expect for describing a cuprate system, similar to the case of the attractive Hubbard model in the main text, the best estimation of $T_c$ that matches the physical observables (here unattainable) likely comes from the extrapolation that the parameter $b$ is small, driving it to values close to $T_c\simeq 0.3t$. Yet, assuming a typical bandwidth $W \sim 2$eV for the copper-oxide compounds~\cite{Arovas2021} would result in critical temperatures much above room temperature if using a non-interacting bandwidth estimation. There are two explanations, the first is that the bandwidth for single-particle excitations is much larger than that in the presence of finite interactions, and is certainly altered by the large $U/t$ we use. Second, the starting point, the single-layer Hubbard model with this set of parameters (here using $U=6t$ and $t^\prime=-0.2t$), does not quantitatively characterize the cuprates. Lastly, interplane hybridization is known to reduce the pairing fluctuations, and may ultimately drive the thermal transition close to the ones currently observed in copper-based superconductors.

\section{Further remarks}
We have here demonstrated that the sign problem can be used as a tracker to understand criticality of fermionic lattice models in four important examples. That the sign problem depends on the basis in which one describes the system's Hamiltonian is well known. What is less understood is whether the average sign in other local bases can also be used to understand critical properties. Similarly, an in-depth study using other Hubbard-Stratonovich transformations~\cite{batrouni90} would lend insight on the wider generality of our results, on whether scaling behavior can be similarly inferred.

\end{document}